\documentclass[aps,prb,amsmath,twocolumn,superscriptaddress,floatfix,footinbib,showpacs,longbibliography]{revtex4}

\usepackage{graphicx}
\usepackage{epstopdf}

\usepackage[T1]{fontenc}
\usepackage[applemac]{inputenc}
\usepackage{lmodern}
\usepackage[english]{babel}

\usepackage{ae}
\usepackage{units}

\usepackage{amsmath,amssymb,natbib,bm}
\usepackage{psfrag}
\usepackage{subfigure}
\usepackage{amsthm}

\usepackage{fixmath}
\usepackage{booktabs}

\usepackage{slashed}

\usepackage[americaninductors]{circuitikz}
\usepackage{tikz}
\usetikzlibrary{arrows}

\usepackage{url}

\usepackage[colorlinks]{hyperref}

\usepackage{tabularx}

\hypersetup{%
        plainpages=true,
        breaklinks=true,
        hypertexnames=false,
        pageanchor=true,
        colorlinks=true,
        linkcolor={blue},
        citecolor={magenta},
        urlcolor={blue},
        anchorcolor={black}
      }

\newcommand{\figref}[1]{\mbox{Fig.~\ref{#1}}}

\renewcommand{\eqref}[1]{\mbox{Eq.~(\ref{#1})}}

\newcommand{\ket}[1]{|#1\rangle}

\newcommand{\abs}[1]{\left|#1\right|}

\newcommand{\be}{\begin{equation}}
\newcommand{\ee}{\end{equation}}
\newcommand{\bea}{\begin{eqnarray}}
\newcommand{\eea}{\end{eqnarray}}

\begin{document}

\title{Mirror, mirror: Landau-Zener-St\"uckelberg-Majorana interferometry of a superconducting qubit in front of a mirror}


\author{P.~Y.~Wen}
\affiliation{Department of Physics, National Tsing Hua University, Hsinchu 30013, Taiwan}

\author{O.~V.~Ivakhnenko}
\affiliation{B.~Verkin Institute for Low Temperature Physics and Engineering, Kharkov 61103, Ukraine}
\affiliation{Theoretical Quantum Physics Laboratory, Cluster for Pioneering
	Research, RIKEN, Wako-shi, Saitama 351-0198, Japan}

\author{M.~A.~Nakonechnyi}
\affiliation{B.~Verkin Institute for Low Temperature Physics and Engineering, Kharkov 61103, Ukraine}

\author{B.~Suri}
\affiliation{Department of Instrumentation and Applied Physics, Indian Institute of Science, Bengaluru 560012, India}

\author{J.-J.~Lin}
\affiliation{Department of Physics, National Tsing Hua University, Hsinchu 30013, Taiwan}

\author{W.-J.~Lin}
\affiliation{Department of Physics, National Taiwan University, Taipei 10617, Taiwan}

\author{J.~C.~Chen}
\affiliation{Department of Physics, National Tsing Hua University, Hsinchu 30013, Taiwan}

\author{S.~N.~Shevchenko}
\email[e-mail:]{sshevchenko@ilt.kharkov.ua}
\affiliation{B.~Verkin Institute for Low Temperature Physics and Engineering, Kharkov 61103, Ukraine}
\affiliation{Theoretical Quantum Physics Laboratory, Cluster for Pioneering
	Research, RIKEN, Wako-shi, Saitama 351-0198, Japan}		
\affiliation{V.~N.~Karazin Kharkiv National University, Kharkov 61022, Ukraine}

\author{Franco~Nori}
\affiliation{Theoretical Quantum Physics Laboratory, Cluster for Pioneering
	Research, RIKEN, Wako-shi, Saitama 351-0198, Japan}
\affiliation{Department of Physics, The University of Michigan, Ann Arbor, MI 48109-1040, USA}

\author{I.-C.~Hoi}
\email[e-mail:]{ichoi@phys.nthu.edu.tw}
\affiliation{Department of Physics, National Tsing Hua University, Hsinchu 30013, Taiwan}
\affiliation{Center for Quantum Technology, National Tsing Hua University, Hsinchu 30013, Taiwan}

\date{\today}
\begin{abstract}
We investigate the Landau-Zener-St\"uckelberg-Majorana interferometry of a superconducting qubit in a semi-infinite transmission line terminated by a mirror. The transmon-type qubit is at the node of the resonant electromagnetic (EM) field, \textquotedblleft hiding\textquotedblright\ from the EM field. \textquotedblleft Mirror, mirror\textquotedblright\ briefly describes this system, because the qubit acts as another mirror. We modulate the resonant frequency of the qubit by applying a sinusoidal flux pump. We perform spectroscopy by measuring the reflection coefficient of a weak probe in the system. Remarkable interference patterns emerge in the spectrum, which can be interpreted as multi-photon resonances in the dressed qubit. Our calculations agree well with the experiments. 

\end{abstract}


\maketitle

\section{Introduction}

\begin{figure*}
	\includegraphics[width=0.8 \linewidth]{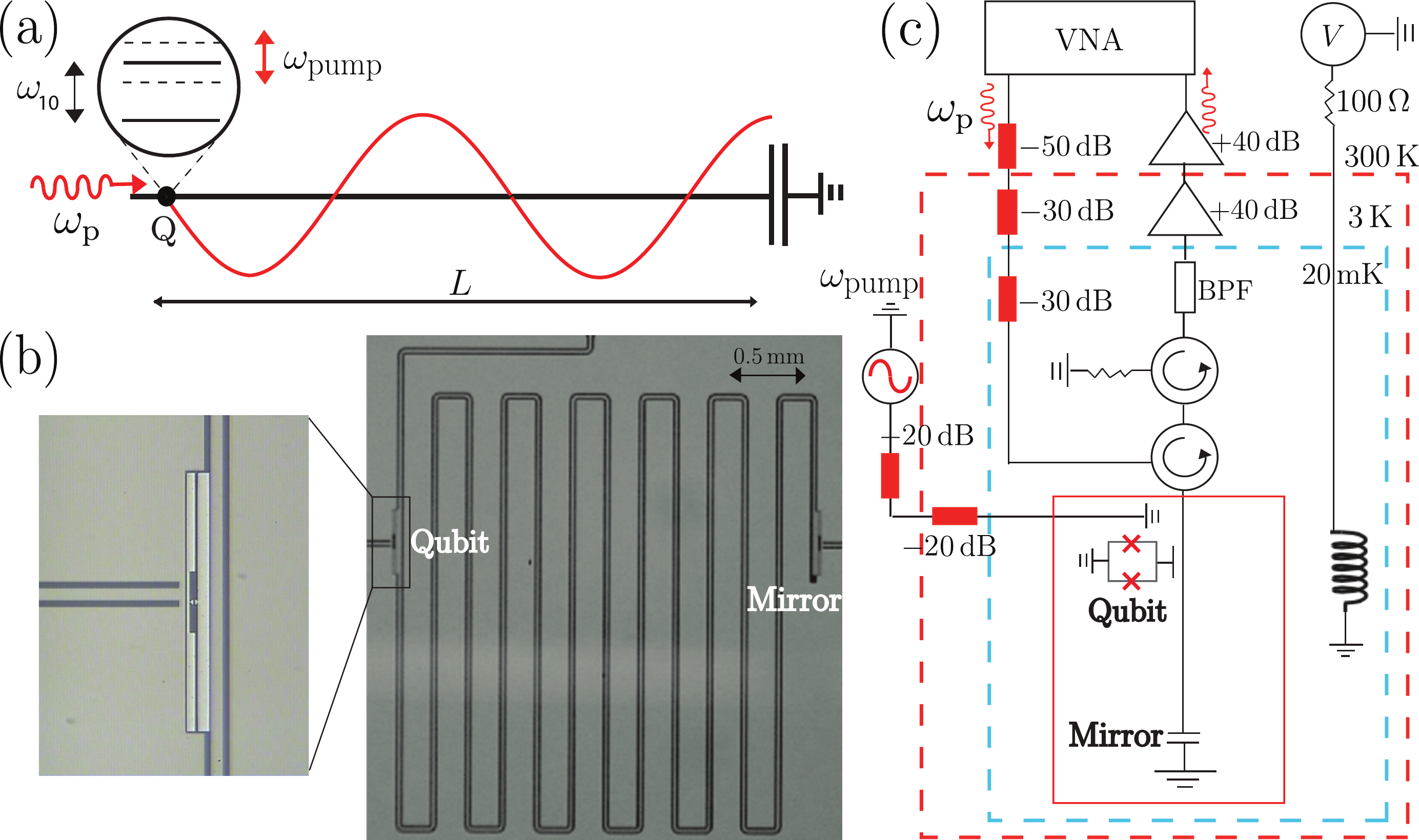}
	\caption{ The experimental setup and device (a) A conceptual sketch of the device showing the electromagnetic mode (red curve) in the transmission line. The qubit is located at the node of the resonant mode of the EM field, hiding from the EM field. The qubit is subjected to a sinusoidal drive along the transmission line with frequency $\omega_{\rm p}$, and a flux pump $\omega_{\rm pump}$.
		(b) A photo of the device. The qubit (shown in the zoom-in on the left; the two long bright parts form the qubit capacitance and the gap in the middle between them is bridged by two Josephson junctions forming a SQUID loop) is placed $L\simeq\unit[33]{mm}$ from another qubit, which sits at the end of the transmission line (i.e., at the mirror). Note that the qubit at the end of the transmission line is not in use in this work, because it is far detuned. The characteristic impedance of the transmission line is $Z_0 \simeq \unit[50]{\Omega}$. By tuning the qubit transition frequency $\omega_{10}(\Phi)$, we tune the qubit to the node of the EM field. 
		(c) A sketch of the setup for the experiment. The qubit frequency $\omega_{10}$ can be tuned by a global magnetic field from a superconducting coil controlled by a dc voltage $V$. For measurements, a coherent signal at frequency $\omega_{\rm p}$ is generated by a vector network analyzer (VNA) at room temperature and fed through attenuators (red squares) to the sample, which sits in a cryostat cooled at $\unit[20]{mK}$ to avoid thermal fluctuations affecting the experiment. The reflected signal passes a bandpass filter (BPF) and amplifiers, and is then measured with the VNA. Here $\omega_{\rm pump}$ relates to the flux pump to modulate the transition frequency of the qubit. 
		\label{fig:Device}}
\end{figure*}

In recent years, superconducting artificial atoms~\cite{Gu2017} coupled to open transmission-line waveguide have been a fast growing field, called waveguide Quantum Electrodynamics (w-QED), which provides a unique platform to investigate atom-light interactions. The uniqueness of w-QED, as compared to conventional cavity QED, is that atoms are coupled to continuum modes of the electromagnetic (EM) field in the waveguide. Exciting problems in w-QED include: resonance fluorescence of an artificial atom~\cite{Astafiev2010}, photon-mediated interactions between distant artificial atoms~\cite{Lalumiere13}, atom in front of a mirror~\cite{Hoi2015}, time dynamics in atom-like mirrors~\cite{Mirhosseini2019}, photon routing~\cite{Hoi2011}, generation of non-classical microwaves~\cite{Hoi2012}, cross-Kerr effect~\cite{Hoi2013a}, amplification without population inversion~\cite{wen18}, collective Lamb shift between two distant artificial atoms~\cite{Wen2019}, ultra strong coupling~\cite{FornDiaz2017}, quantum rifling~\cite{Szombati19}, probabilistic motional averaging~\cite{Karpov2020}, and the dynamical Casimir effect~\cite{Johansson2009, Johansson2010, Wilson2011, Lahteenmaki2013}.

When a two-level system is driven back and forth around its resonance frequency, it will produce Landau-Zener-St\"uckelberg-Majorana (LZSM) interference.  LZSM interferometry~\cite{Oliver2005, Sillanpaa2006, Shevchenko2010} has been studied in atomic systems~\cite{Ditzhuijzen}, quantum dots~\cite{Bogan, Mi2018}, charge and spin qubits~\cite{Ono19, Otxoa2019} and superconducting qubit in cavity~\cite{Li13, Silveri15, Pan17, Bera2020}, among others. However, the effect of LZSM has not been explored with a single artificial atom in front of a mirror, where the artificial atom is coupled to a continuum of modes of the EM field in the transmission-line waveguide, and the atom interferes with its mirror image, as in Refs.~[\onlinecite{Hoi2015, Eschner}]. 

LZSM interferometry is important for both system description and control~\cite{Gorelik1998, Wu19}. However, for this to be realized, one needs to have the avoided energy-level crossing in the spectrum as a function of a controlling parameter. One example of systems without this are transmon-type superconducting qubits, where the energy levels are almost independent of the gate voltage. The way to fix this was studied in Ref.~[\onlinecite{Gong16}], which studied the qubit by chirping the microwave frequency, which results in dressed states with avoided-level crossing. In this work, we study a transmon qubit driven by two fields (see also Ref.~[\onlinecite{Satanin2014}]). One of these dresses the qubit and creates the spectrum with the avoided-level crossing, while the other one makes the system periodically pass around the avoided-level point. This allows to study LZSM interferometry in a qubit placed in front of a mirror. \textquotedblleft Mirror, mirror\textquotedblright\ briefly describes this system, because the qubit acts as another mirror.

\section{Superconducting qubit in front of a mirror}

In this work, we investigate the LZSM interferometry of a superconducting qubit in a semi-infinite transmission line, at a distance terminated by a mirror. In particular, the qubit is located at the node of the resonant EM field, where it is hiding from the EM field. We then modulate the resonant frequency of the qubit by applying a sinusoidal wave through an on-chip flux pump. In addition, the qubit is also coupled to the microwave probe signal applied to the transmission line. We then perform the spectroscopy of the system by applying a weak probe field along the transmission line and measure the reflection coefficient. Interesting interference patterns emerge in the spectrum, which can be explained by multi-photon resonances in the dressed qubit. New features appear, as compared to conventional LZSM interference; for example, now the zero-order Rabi sideband vanishes (see also Ref.~[\onlinecite{Giavaras2020}]).

Figure 1 shows (a) a sketch of the device, (b) the image of the device, and (c) the measurement setup. A transmon qubit~\cite{You2006, You2007, Kockum2019} is embedded in a semi-1D transmission line with characteristic impedance $Z_0 \simeq \unit[50]{\Omega}$, with the ground state $\ket{0}$ and the excited state $\ket{1}$. The $\ket{0} \leftrightarrow \ket{1}$ transition energy is 

\begin{equation}
\hbar \omega_{10}(\Phi) \approx \sqrt{8 E_J(\Phi) E_C} - E_C,
\label{omega10}
\end{equation}%
which is determined by the single-electron charging energy $E_C = e^2 / 2C_\Sigma$, where $C_\Sigma$ is the total capacitance of the qubit, and the flux-dependent Josephson energy $E_J(\Phi) = E_{J,\max} \abs{\cos(\pi \Phi / \Phi_0)}$; $\Phi_0 = h / 2e$ is the magnetic flux quantum. The ratio $E_C/E_J$ determines the anharmonicity of the qubit. 

In \figref{fig:Device}(c), a probe field of frequency $\omega_{\rm p}$ is fed into the transmission line. The pump field of frequency $\omega_{\rm pump}$ is applied to the on-chip flux line, sinusoidally modulating the transition frequency of the qubit. The key parameters are summarized in Table~\ref{Table1}.

\begin{table}[tbp]
	\begin{tabularx}{\linewidth}{|>{\hsize=0.4\hsize}X|>{\hsize=1.8\hsize}X|>{\hsize=0.8\hsize}X|}
		\hline
		
		\centering Value & \centering Description & Range \\ \hline
		
			\centering $ \omega _{\mathrm{10}}$ & qubit frequency, $\omega _{10}=\omega _{10}(V)$ & $ \simeq \omega _{\mathrm{node}}$ \\ \hline
		
			 \centering $\delta$ & pump amplitude; $\delta = \delta(P_{\mathrm{pump}})$ &  $ \sim 0.1$ GHz$ \cdot  2\pi$ \\ \hline
			 
			 \centering $\omega_{\mathrm{pump}}$ & pump frequency &  $<0.1$ GHz$ \cdot  2\pi$ \\ \hline
			 
			 \centering $\omega_{\mathrm{p}}$ & probe frequency & $ \simeq \omega _{\mathrm{node}}$ \\ \hline		
		
	\end{tabularx}
\caption{Table of controllable parameters. Here $\omega_{\rm node} = \unit[4.75]{GHz} \! \cdot \! 2\pi$.}

\label{Table1}
\end{table}

\section{Theoretical description}

Let us now consider the qubit Hamiltonian, with details presented in Appendix A. Thanks to the mirror, the transmission-line voltage
at the point of coupling the qubit, $x=L$, is proportional to $\cos
\left( \omega _{\mathrm{p}}L\right/v) $. When this factor is zero,
this gives the node frequency $\omega _{\mathrm{node}}$, with $\cos \left( \omega _{\mathrm{node}}L/v \right) =0$. For small frequency offset, 
\begin{equation}
\Delta \omega=\omega _{\mathrm{p}}-\omega _{\mathrm{node}}\ll \omega _{\mathrm{p}},
\label{Deltaomega}
\end{equation}%
we can expand the cosine into series; then instead of $\cos \left( \omega _{\mathrm{p}}L /v \right) $, we have $%
\Delta \omega /\omega _{\mathrm{node}}$. This means that at $\Delta \omega
=0$ the qubit is \textquotedblleft hidden\textquotedblright\ or
\textquotedblleft decoupled\textquotedblright\ from the transmission line.
So, we have the Hamiltonian 
\begin{equation}
H=-\frac{B_{z}}{2}\sigma _{z}-\frac{B_{x}}{2}\sigma _{x},
\label{Hamiltonian}
\end{equation}%
of which the diagonal part is the energy-level modulation 
\begin{equation}
B_{z}/\hbar =\omega _{10}+\delta \sin \omega _{\mathrm{pump}}t,  \label{Bz}
\end{equation}%
and the off-diagonal part describes the coupling to the probe signal%
\begin{equation}
B_{x}/\hbar =G\sin \omega _{\mathrm{p}}t.  \label{Bx}
\end{equation}%
Importantly, here the coupling constant $G$ is proportional to the frequency
offset $\Delta \omega$,
\begin{equation}
G=G_{0}\frac{\Delta \omega }{\omega _{\mathrm{node}}},
\end{equation}
and $G_{0}$ is proportional to the probe signal amplitude.



The observable value is the reflection coefficient $r$, namely its deviation from 1. The impact of the qubit is in suppressing $r$. Thus, following Refs.~[\onlinecite{Li13, Silveri15, Wu19}], we associate the reflection coefficient to the qubit upper-level occupation probability $P_1$. This will be analyzed in the following Section.

\section{LZSM interference meets multi-photon excitations}

\begin{figure}
	\includegraphics[width=0.8 \linewidth]{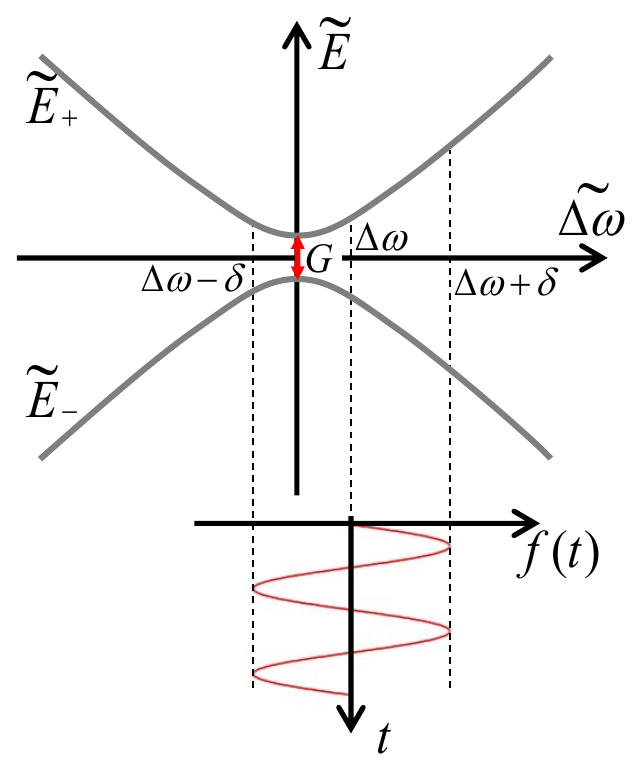}
	\caption{Dressed-state energy levels $\widetilde{E}_{\pm }$ as a function of the bias 
		$\widetilde{\Delta \omega }$. Sinusoidal driving $f(t)$, shown at the
		bottom, makes the system evolve periodically between $\Delta \omega -\delta $
		and $\Delta \omega +\delta $. 
		For small energy-level splitting $G$ and strong driving (with large amplitude $\delta $), the resonant excitation of the system can equivalently be described either in terms of sequential LZSM transitions or in terms of multi-photon excitations, at $\Delta \omega =k\omega _{\mathrm{pump}}$.
		\label{FigN}}
\end{figure}

To remove the fast driving from the Hamiltonian, we perform the unitary
transformation $U=\exp \left( -i\omega _{\mathrm{p}}\sigma _{z}t/2\right) $
and the rotating-wave approximation \cite{Silveri15, Ono2020}. The new
Hamiltonian takes the form%
\begin{equation}
H_{1}=-\!\frac{\hbar \widetilde{\Delta \omega }}{2}\sigma _{z}+\frac{\hbar G%
}{2}\sigma _{x},  \label{H_in_RWA}
\end{equation}%
where%
\begin{eqnarray}
\widetilde{\Delta \omega } &=&\Delta \omega +f(t), \\
\Delta \omega  &=&\omega _{\mathrm{p}}-\omega _{\mathrm{10}}, \\
f(t) &=&\delta \sin \omega _{\mathrm{pump}}t.
\end{eqnarray}%
The eigenstates of the Hamiltonian $H_{1}$ can be called dressed states,
since they incorporate the microwave driving into the qubit-like Hamiltonian
(\ref{H_in_RWA}). These dressed states have energy levels derived from
Eq.~(\ref{H_in_RWA}):%
\begin{equation}
\widetilde{E}_{\pm }=\!\pm \frac{\hbar }{2}\sqrt{G^{2}+\widetilde{\Delta
		\omega }^{2}}.
\end{equation}%
These are illustrated in \figref{FigN}.

The dynamics of the system, implied in \figref{FigN}, can be conveniently described in
terms of LZSM interference~[\onlinecite{Shevchenko2010}]. Being driven
by the slow signal $f(t)$, the system periodically evolves around $\widetilde{%
	\Delta \omega }=\Delta \omega $. When the system comes around the
avoided-level crossing, where the energy difference  $\Delta \widetilde{E}=%
\widetilde{E}_{+}-\widetilde{E}_{-}$ has a minimum equal to $\hbar G$, the system
can be partially excited by means of LZSM transitions. Periodically
traversing the avoided-level crossing produces
interference. This is described by the phase, accumulated by the vector-state
during one period%
\begin{equation}
\zeta =\int\limits_{0}^{2\pi /\omega _{\mathrm{pump}}}dt\Delta \widetilde{E}%
/\hbar \approx \Delta \omega \frac{2\pi }{\omega _{\mathrm{pump}}}.
\end{equation}%
Here, the approximation means the assumption of small splitting and strong
driving: $G\ll \delta $. When the phase $\zeta $ equals to $2\pi k$, with an
integer $k$, the system is resonantly excited. Then, from $\zeta =2\pi k$
the resonance condition becomes%
\begin{equation}
\Delta \omega =k\omega _{\mathrm{pump}}.  \label{multi}
\end{equation}%
This can be interpreted as a multi-photon excitation, meaning that the
system is resonantly excited when the dressed energy gap $\hbar \Delta \omega 
$ equals to the energy of $k$ photons, $k\hbar \omega _{\mathrm{pump}}$.

Quantitatively, from the stationary solution of the Bloch equations with the
Hamiltonian $H_{1}$, in the rotating-wave approximation, for
the upper-level occupation probability we have (see e.g. Refs.~[%
\onlinecite{Silveri2017, Ivakhnenko18}]):

\begin{equation}
P_{1}=\frac{1}{2}\sum\limits_{k=-\infty }^{\infty }\frac{G_{k}^{2}}{%
	G_{k}^{2}+\left[ \Delta \omega -k\omega _{\mathrm{pump}}\right] ^{2}\frac{%
		\Gamma _{1}}{\Gamma _{2}}+\Gamma _{1}\Gamma _{2}},  \label{P1}
\end{equation}%
where the renormalized driving amplitude 
\begin{equation}
G_{k}=GJ_{k}(\delta /\omega _{
	\mathrm{pump}}),  \label{Gk}
\end{equation}
follows the oscillating Bessel function $J_{k}$ of the
first kind; $\Gamma _{1}$ and $\Gamma _{2}=\Gamma _{1}/2+\Gamma _{\phi }$
are the relaxation and decoherence rates, with the pure dephasing rate $%
\Gamma _{\phi }$ being much smaller than $\Gamma _{1}$. One can
see that the maximum of $P_{1}$ is indeed defined by the condition (\ref%
{multi}). With this formula, Eq.~(\ref{P1}), we plot
theoretical graphs in Figs.~3,~4, and~5. For details see Appendix B. 

\begin{figure}
	\includegraphics[width=\linewidth]{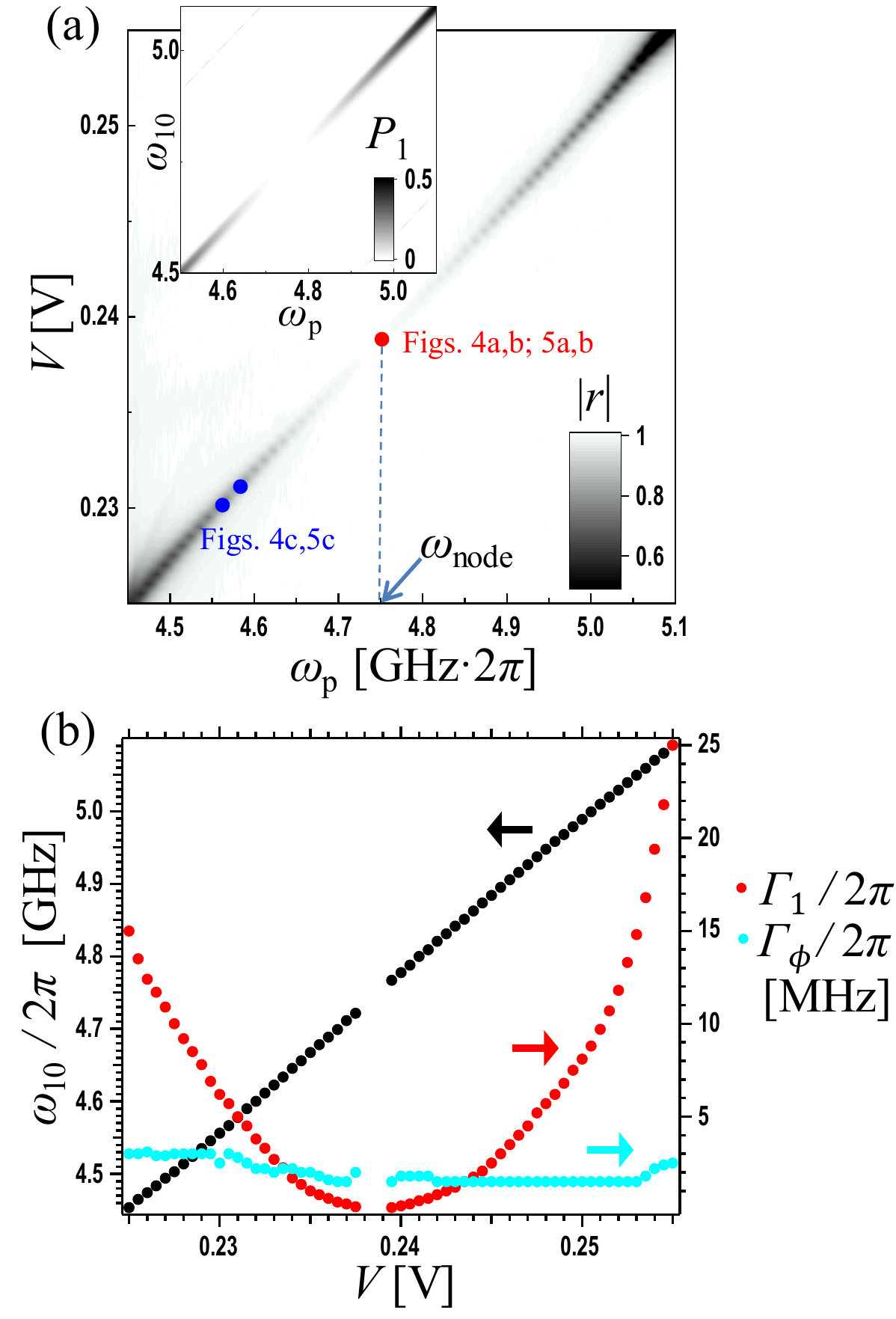}
	\caption{Spectroscopy of the system. (a) Amplitude of the reflection coefficient $\abs{r}$ at a coherent probe power of $\unit[-130]{dBm}$ as a function of probe frequency $\omega_{\rm p}$ and qubit frequency $\omega_{10}$ (controlled by the voltage $V$). The spectroscopy shows how the response disappears when the qubit ends up at a node for the EM field around $\omega_\mathrm{p}=\omega_\mathrm{node}\simeq\unit[4.75]{GHz}\! \cdot \! 2\pi$, denoted by the dashed line marker. The red dot indicates the qubit bias point in Fig.~4a,b and Fig.~5a,b. The blue dots correspond to the qubit bias point in Fig.~4c and Fig.~5c, respectively. The inset shows the calculated qubit upper-level occupation probability $P_{1}$ as a function of the probe frequency $\omega_\mathrm{p}$ and the qubit frequency $\omega_{10}$ (both in GHz$ \cdot 2\pi$). (b) For a given bias voltage in (a), it is a Lorentzian dip (data not shown) indicating that the qubit reflects the resonant probe field. The qubit acts as a mirror, reflect the resonant field. We extract the resonant frequency $\omega_{10}$, relaxation rate $\Gamma_{1}$, and pure dephasing rate $\Gamma_{\phi}$ following Ref.~[\onlinecite{HoiPhD}], taking into account the effect of the Rabi frequency. In particular, we convert the probe power to Rabi frequency at a resonant frequency of $\unit[5.05]{GHz}$ using calibration data in Ref.~[\onlinecite{Wen2019}] (same sample, but in a different cool down). Moreover, the Rabi frequency at each resonant frequency is given by Eq.~(4). Finally, we see $\omega_{10} \propto V $; however, in general, for the wide range of voltage bias, this is not the case; see details of the flux dependence in Ref.~[\onlinecite{Wen2019}].
		\label{fig:Flux}}
\end{figure}

\begin{figure*}
	\includegraphics[width=1 \linewidth]{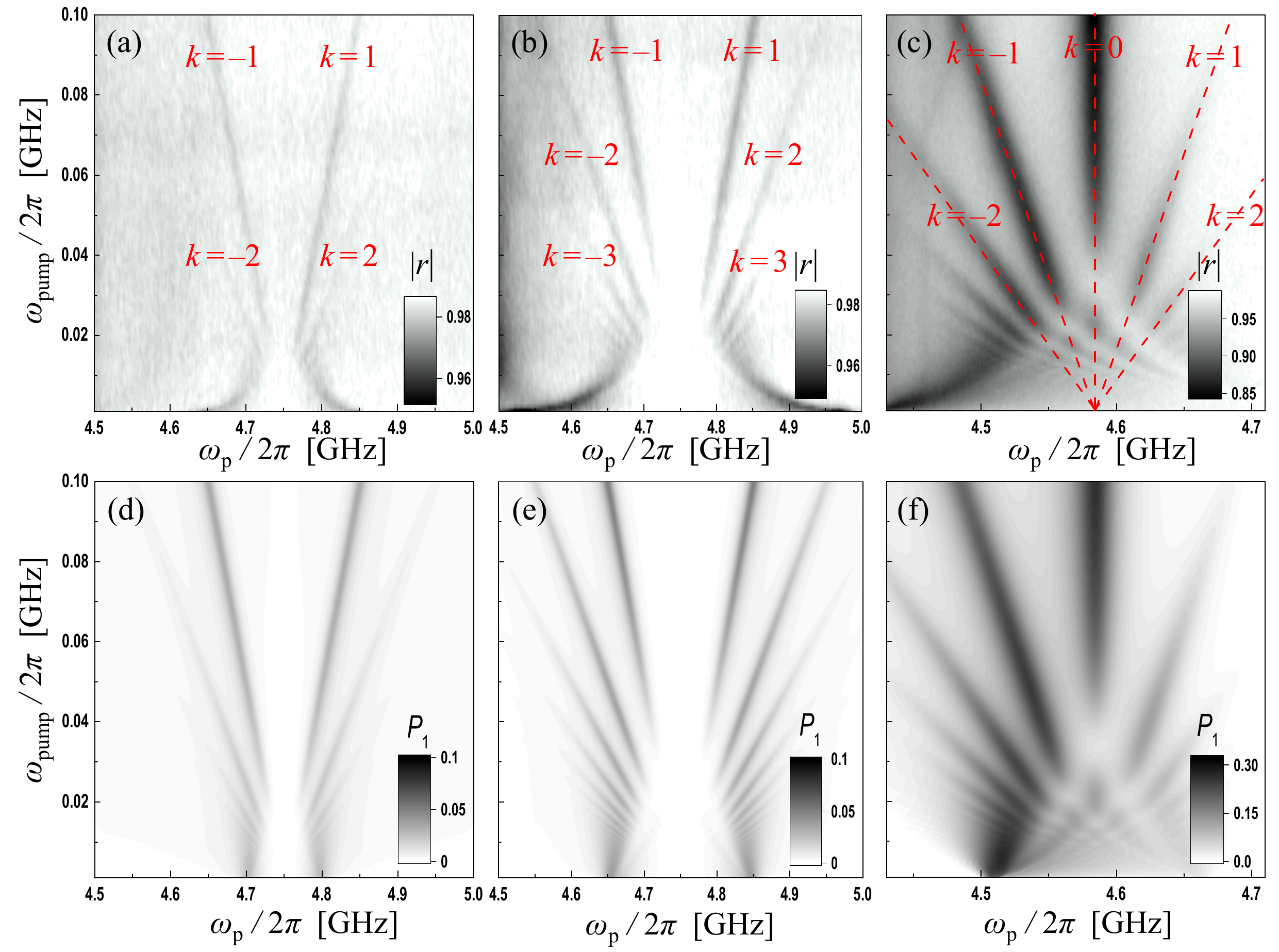}
	\caption{Sinusoidal modulation of the qubit by flux pumping, with the resonance frequency at a fixed pump power $P_{\rm pump}$. Amplitude reflection coefficient $\abs{r}$ for a weak coherent probe as a function of the probe frequency $\omega_{\rm p}$ and pump frequency $\omega_{\rm pump}$;  (a,b,c) are experimental data, while (d,e,f) are our theoretical calculations. (a) Qubit biased at $\omega_{\rm node}=\unit[4.75]{GHz} \! \cdot \! 2\pi$, the flux pump power is fixed at $P_{\rm pump}=\unit[-45]{dBm}$. (b) Qubit biased at the node around $\unit[4.75]{GHz}$ with $P_{\rm pump}=\unit[-38]{dBm}$. (c) Qubit biased at $\unit[4.58]{GHz}$, red detuned from the node, and $P_{\rm pump}=\unit[-38]{dBm}$. Note that the $k$-dependent multi-photon resonances emerge: $\Delta\omega=k\omega_{\rm pump}$, which are shown by the inclined red dashed lines. In (b), we can see Rabi sidebands for $k$ from $-4$ to 4. In (a) and (b), the $k=0$ Rabi sideband disappears whereas in (c) the $k=0$ Rabi sideband appears. For (a) and (b), the positive $k$ and negative $k$ fringes are symmetric, whereas, in (c) the interference fringes are not symmetric along $k=0$. As the fringes approach the node regime, near $\unit[4.75]{GHz}$, they become weaker. In (d-f) we show the respective calculated qubit upper-level occupation probabilities $P_{1}$. Experimentally, the pump power reached at the sample is slightly frequency dependent. The low frequency flux pumping drive (say \unit[1]{MHz}) has about one dB higher power than the high frequency flux pumping drive (say \unit[20]{MHz}). Therefore, we see that the $\omega_{\rm pump}$ at low frequency has more shift $\delta$ in the data.  
		\label{fig:PumpFrequencyDependent}}
\end{figure*}

\begin{figure*}
	\includegraphics[width=1 \linewidth]{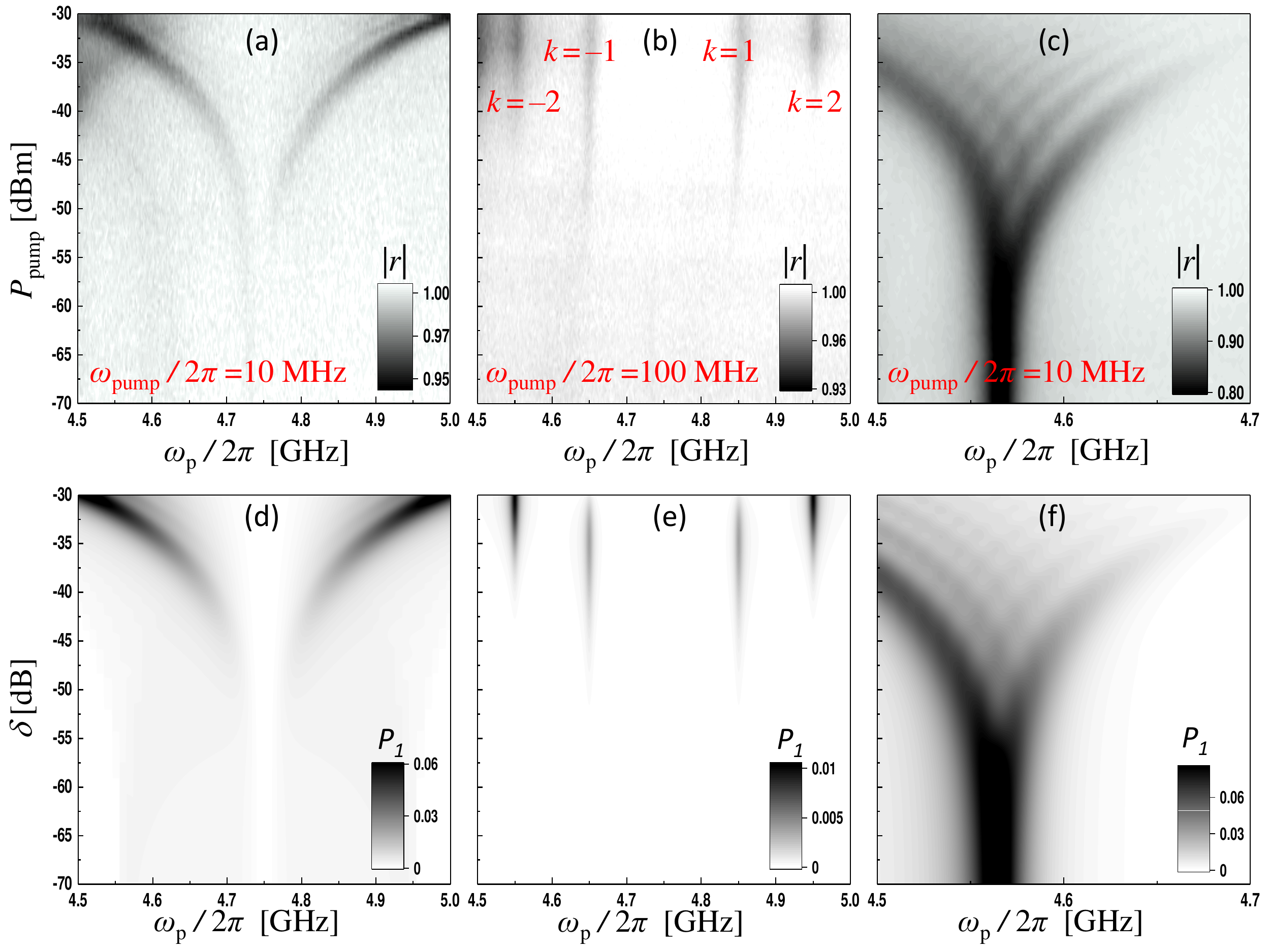}
	\caption{Sinusoidal modulation of the qubit by flux pumping at the resonance frequency at a fixed pump frequency $\omega_{\rm pump}$. The plots show the amplitude of the reflection coefficient $\abs{r}$ for a weak coherent probe versus the probe frequency $\omega_{\rm p}$ and pump power $P_{\rm pump}$. (a,b,c) correspond to experiments and (d,e,f) to theory. (a) and (b) are at the node around $\unit[4.75]{GHz}$ with $\omega_{\rm pump}/2\pi=\unit[10]{MHz}$ and $\unit[100]{MHz}$, respectively. Note the Rabi sidebands at $\Delta\omega=k\omega_{\rm pump}$, with $k=-2, -1, +1, +2$. The higher the pump power, the more resolved sidebands are visible. Note the onset of the Rabi sidebands for $k=\pm1$ and $k=\pm2$, for $\unit[-45]{dBm}$ and $\unit[-35]{dBm}$, respectively. (c) red detuned from the node at $\unit[4.56]{GHz}$. In (a,c), we observe Rabi-like splittings as the pump power increases. In (a), we see a symmetric splitting, whereas in (c) there are several asymmetric splittings. In (c), as the fringes approach $\omega_{\rm node}$, they become weaker. In (d-f) we show the respective calculated qubit upper-level occupation probabilities $P_{1}$. 
		\label{fig:PumpPowerDependent}}
\end{figure*}

\section{Measurements}

We first perform single-tone spectroscopy of the qubit-mirror system. In \figref{fig:Flux}, the resonant frequency of the qubit is tuned by voltage. As the voltage increases, the linewidth of the qubit decreases from a finite linewidth to zero, and then increases back to a finite linewidth. At the frequency where the linewidth vanishes, around $\omega_{10}=\omega_{\rm node}\simeq\unit[4.75]{GHz} \! \cdot \! 2\pi$, the qubit is located at the node of the EM field, as indicated by the vertical dashed line, where it is hiding from the EM field. 

By using two-tone spectroscopy~\cite{Wen2019}, we know that $E_C/h\simeq\unit[324]{MHz}$. For $\omega_{10}/2\pi=\unit[4.75]{GHz}$, the corresponding Josephson energy is $E_J/h\simeq\unit[9.9]{GHz}$. 

After the basic characterization of the system, we want to study the spectrum as a function of the following parameters: qubit frequency $\omega_{10}$, pump amplitude $\delta$, pump frequency $\omega_{\rm pump}$, and probe frequency $\omega_{\rm p}$. For spectroscopy, we always use the probe power of $\unit[-130]{dBm}$ in experiments. 

To start with, we set the qubit frequency corresponding to the node as a working point, where $\omega_{10}=\omega_{\rm node}$. We then apply a sinusoidal flux pump at a fixed power to the qubit, and sweep the pump frequency from $\unit[1]{MHz}$ to $\unit[100]{MHz}$.  At the same time, we perform the spectroscopy of the system using a weak field, with the frequency $\omega_{\rm p}$ near the qubit frequency. 

We show the amplitude reflection coefficient $\abs{r}$ in \figref{fig:PumpFrequencyDependent}(a,b) as a function of $\omega_{\rm pump}$ and  $\omega_{\rm p}$ in (a) for $P_{\rm pump}=\unit[-45]{dBm}$, in (b) for $P_{\rm pump}=\unit[-38]{dBm}$. We observed LZSM interference fringes. These interference fringes can be interpreted as multi-photon resonances in the dressed qubit. Multi-photon resonances appear at $\omega_{\rm p}=\omega_{10}\pm k\, \omega_{\rm pump}$, where $k$ is the order, as indicated in the figures. The zero order, where $k=0$, is missing, which is a key feature here, different from conventional LZSM interference fringes. In \figref{fig:PumpFrequencyDependent}(b), we can clearly see the order $k$ up to $\pm4$. We increase the pump power in \figref{fig:PumpFrequencyDependent} from $\unit[-45]{dBm}$ in (a) to $\unit[-38]{dBm}$ in (b), and the gap between negative $k$ and positive $k$ fringes becomes wider. Indeed, the stronger the pump power, the wider they separate (data not shown). The power in \figref{fig:PumpFrequencyDependent} (a) and (b) differs by $\unit[7]{dB}$, meaning that their pump amplitude $\delta$ differs by a factor of 2.2. This is also what happens in the theory calculation plots. In this sense, this separation can be used to calibrate the pump power. 

In\figref{fig:PumpFrequencyDependent}(c), we bias the qubit at around $\unit[4.58]{GHz}$, red detuned from the node, and we then see asymmetric interference fringes. At this bias point, when the qubit is pumped in the negative part of the sinusoidal, the qubit is pumped toward the larger linewidth regime, see \figref{fig:Flux}. However, when the qubit is pumped in the positive part of the sinusoidal, the qubit is pumped towards the zero-linewidth regime; therefore, we can see that the interference fringes vanish near $\omega_{\rm node}$. In addition, in contrast to Fig.~4(a) and (b), the $k=0$ zero Rabi sideband appears in Fig.~4(c) because there the qubit is biased away from the node.

Next, we keep the pump frequency $\omega_{\rm pump}$ constant. We change the power of the pump $P_{\rm pump}$ from $\unit[-70]{dBm}$ to $\unit[-30]{dBm}$, and probe the system with a weak field near the resonance frequency of the qubit. In \figref{fig:PumpPowerDependent}, we show in (a) and (b) $\omega_{\rm pump} / 2\pi =\unit[10]{MHz}$ and $\unit[100]{MHz}$ for the qubit frequency at the node $\omega_{\rm node}$, and $\omega_{\rm pump} / 2\pi =\unit[10]{MHz}$ for the qubit frequency red detuned from $\omega_{\rm node}$ in (c). These plots show the amplitude reflection coefficient $\abs{r}$ as a function of the probe frequency $\omega_{\rm p}$ and the pump power $P_{\rm pump}$. In \figref{fig:PumpPowerDependent}(a,b), when the flux pump is weak (this corresponds to a small change of the qubit resonance frequency) there are no interference fringes in the interference pattern. This can be explained using \figref{fig:Flux}, where the \textquotedblleft node regime\textquotedblright\ corresponds from $\unit[4.7]{GHz}$ to $\unit[4.8]{GHz}$, and there is no response for a weak flux pump. When the pump power increases, this corresponds to larger changes of the resonance frequency, and we see the Rabi-splitting-like behavior in (a). In (b), Rabi sidebands at $k=-2, -1, +1, +2$ appear. These match the condition $\Delta\omega=k\,\omega_{\rm pump}$. The higher the pump power, the more resolved the sideband $k$ becomes. In \figref{fig:PumpPowerDependent}(c), when we bias the qubit frequency away from $\omega_{\rm node}$, the interference fringes become weaker as the probe frequency approaches the node frequency at $\unit[4.75]{GHz}$.

\section{Conclusions}

In conclusion, we investigate the LZSM interferometry of a superconducting qubit in a semi-infinite transmission line terminated by a mirror. When the qubit frequency is set to the node of the EM field, after flux pumping the qubit frequency, remarkable interference patterns emerge, which can be interpreted as multi-photon resonances in the dressed qubit. We see multi-photon resonances up to the 4th order. Since the qubit interferes with its mirror image, and the zero-order photon resonance disappears. Such effect would not appear in the case of an infinite transmission line. One of the advantages of this atom-mirror arrangement is that we can effectively manipulate the absorption properties of the two-level atom, providing a novel way to manipulate the quantum states.  

\begin{acknowledgments}


I.-C.H. and J.C.C. would like to thank I. A.~Yu and C.-Y.~Mou for fruitful discussions. S.N.S. is grateful to A.I.~Krivchikov for useful discussions. This work was financially supported by the Center for Quantum Technology from the Featured Areas Research Center Program within the framework of the Higher Education Sprout Project by the Ministry of Education (MOE) in Taiwan. I.-C.H.~acknowledges financial support from the MOST of Taiwan under project 109-2636-M-007-007. J.C.C. acknowledges financial support from the MOST of Taiwan under project 107-2112-M-007-003-MY3. O.V.I., M.A.N., and S.N.S. acknowledge partial support by the Grant of the President of Ukraine (Grant No.~F84/185-2019). O.V.I. and S.N.S. were supported in part by Army Research Office (ARO) (Grant No.~W911NF2010261). F.N. is supported in part by: NTT Research, Army Research Office (ARO) (Grant No. W911NF-18-1-0358), Japan Science and Technology Agency (JST) (via Q-LEAP and the CREST Grant No. JPMJCR1676), Japan Society for the Promotion of Science (JSPS) (via the KAKENHI Grant No. JP20H00134, and the grant JSPS-RFBR Grant No. JPJSBP120194828), and the Grant No. FQXi-IAF19-06 from the Foundational Questions Institute Fund (FQXi), a donor advised fund of the Silicon Valley Community Foundation. 
\end{acknowledgments}

\appendix

\section{Hamiltonian}

\begin{figure}[h]
	\begin{center}
		\includegraphics[width=0.48\textwidth, keepaspectratio]{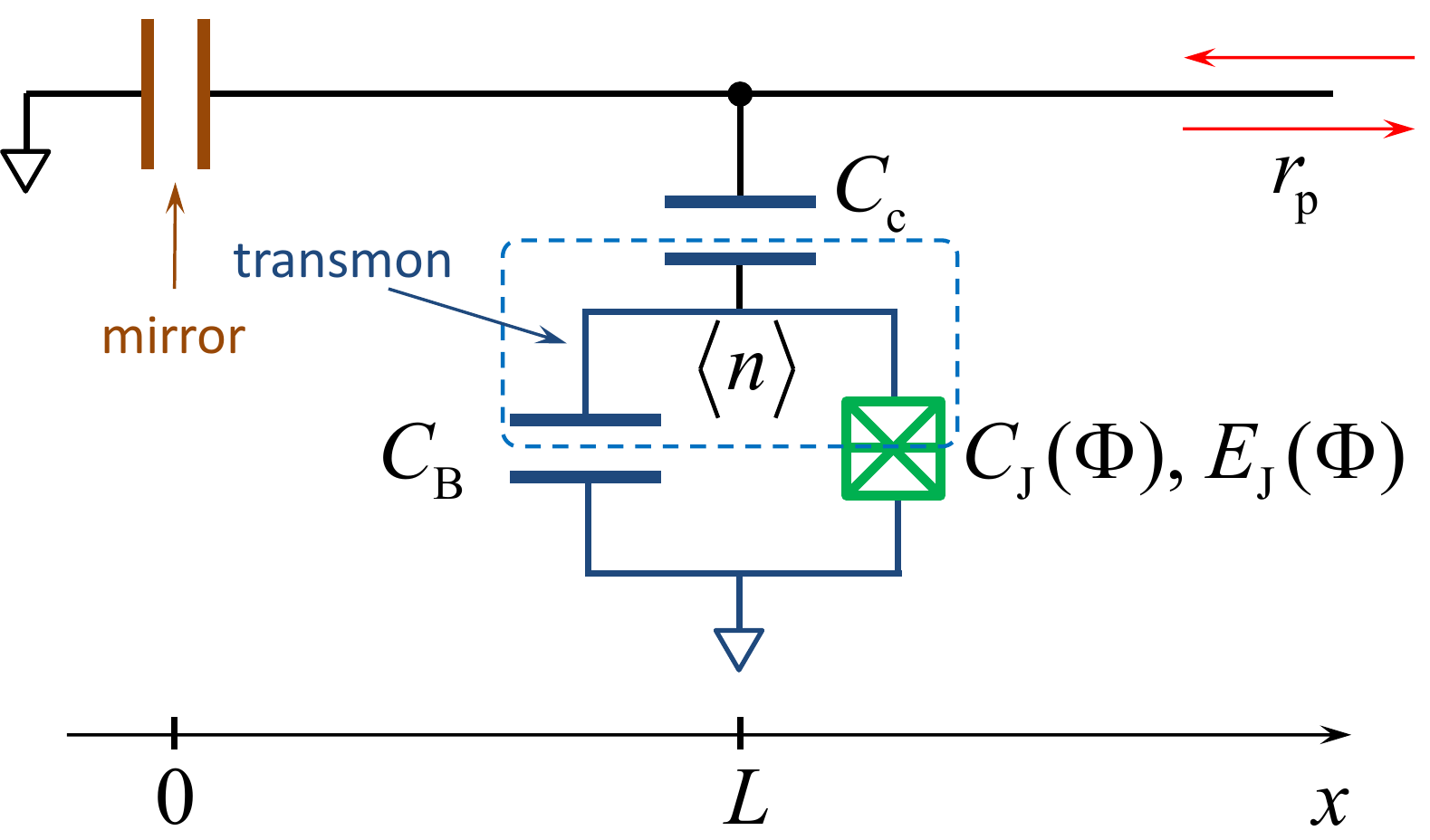}
	\end{center}
	\caption{Schematic diagram of a capacitively shunted charge qubit in front of a mirror. The qubit and the transmission line are coupled through the
		capacitance $C_{\mathrm{c}}$. The transmission line is biased by a probe
		signal and its reflection coefficient $r_{\mathrm{p}}$ is measured. The line
		is terminated by a capacitor, playing the role of a mirror. The transmon
		qubit consists of Josephson junctions, described by the flux-dependent
		capacitance $C_{\mathrm{J}}(\Phi )$, shunted by a capacitance $C_{\mathrm{B}}
		$.}
	\label{Fig:scheme}
\end{figure}

In this Appendix we describe how we obtain the Hamiltonian (\ref{Hamiltonian}%
) for a qubit in front of a mirror, schematically shown in Fig.~\ref{Fig:scheme}, see also Ref.~[\onlinecite{Wendin2017}].

The transmission line is described by the voltage $V(x,t)$ and current $%
I(x,t)$: 
\begin{equation}
V(x,t)=V(x)e^{i\omega _{\mathrm{p}}t},\text{ \ \ }I(x,t)=I(x)e^{i\omega _{%
		\mathrm{p}}t},  \label{V_and_I}
\end{equation}%
with
\begin{eqnarray}
V(x) &=&V_{+}e^{ikx}+V_{-}e^{-ikx},  \label{V(x)} \\
I(x) &=&-\frac{V_{+}}{Z_{0}}e^{ikx}+\frac{V_{-}}{Z_{0}}e^{-ikx},
\label{I(x)}
\end{eqnarray}%
where $k=\omega _{\mathrm{p}}/v$. Thanks to the mirror at $x=0$, we have $%
I(0)=0$, $V_{-}=V_{+}$, and $V(x)=2V_{+}\cos \left( kx\right) $ for $x\in
(0,L)$.

The transmon is described by the number $%
\left\langle n\right\rangle $ of Cooper pairs on it, with the number operator $n$ given by the Pauli
matrix \cite{Koch2007}%
\begin{equation}
n=\left(\frac{E_{\mathrm{J}}}{32E_{\mathrm{C}}}\right)^{1/4}\sigma _{x}.
\end{equation}%
If we take here $\hbar \omega _{10}\approx \sqrt{8E_{\mathrm{C}}E_{\mathrm{J}%
}}$, we have%
\begin{equation}
n =\sqrt{\frac{\hbar \omega _{10}}{E_{\mathrm{C}}}}%
\sigma _{x}.  \label{n}
\end{equation}

\begin{figure*}
	\includegraphics[width=0.7 \linewidth]{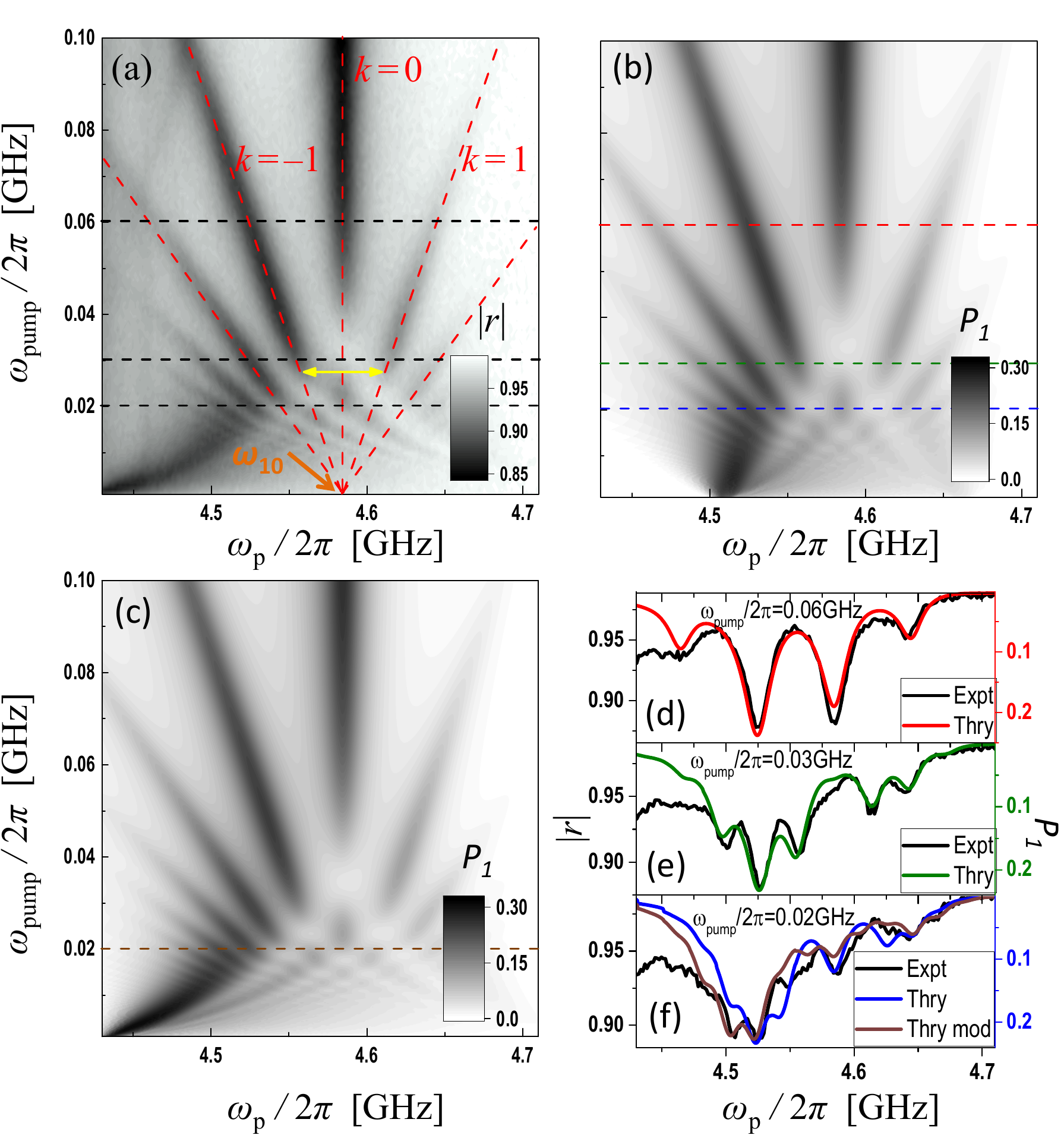}
	\caption{(a) Reflection coefficient as a function of the probe frequency $\omega_{\rm p}$ and the pump frequency $\omega_{\rm pump}$, as in Fig.~\ref{fig:PumpFrequencyDependent}(c). (b) Upper-level occupation probability $P_1$, calculated for the same parameters as Fig.~\ref{fig:PumpFrequencyDependent}(f). (c) Similar interferogram  with taking into account the phenomenological nonlinearity, \eqref{B1}. (d,e,f) The cross-sections taken at different values of the pump frequency, $\omega_{\rm pump}/2\pi=60,30,20$~MHz, respectively. (d) and (e) demonstrate agreement of the theoretical calculations (red and green curves) with the experimental data from (a) (black curves) for high pump frequency, while (f) demonstrates the deviation of the theory with constant $\delta$ (blue curve) from the experiment. The modified theory, with non-linear $\delta^\prime(\omega_{\rm pump})$, the brown curve, is in much better agreement.}
	\label{fig:ExpThrComparison}
\end{figure*}

Then, writing down the charges of the respective capacitor plates, we obtain the
island voltage \cite{Shevchenko18}%
\begin{equation}
V_{\mathrm{I}}=\frac{2e}{C_{\Sigma }} n -\frac{C_{%
		\mathrm{c}}}{C_{\Sigma }}V(L,t),
\end{equation}%
where $C_{\Sigma }=C_{\mathrm{J}}+C_{\mathrm{B}}+C_{\mathrm{c}}$. With this we can write the Hamiltonian of the
transmon qubit coupled to the transmission line, which can be rewritten (omitting c-numbers) as follows 
\begin{eqnarray}
H_{\mathrm{c}} &=&\frac{1}{2}C_{\mathrm{c}}\left[ V(L,t)-V_{\mathrm{I}%
}\right] ^{2}\rightarrow C_{\mathrm{c}}V(L,t)V_{\mathrm{I}}\rightarrow
\label{Hc} \\
&\rightarrow &eV_{+}\frac{C_{\mathrm{c}}}{C_{\Sigma }}\sqrt{\frac{\hbar
		\omega _{10}}{E_{\mathrm{C}}}}\cos \left( \frac{\omega _{\mathrm{p}}}{v}%
L\right) \sin \left( \omega _{\mathrm{p}}t\right) \sigma _{x}.  \notag
\end{eqnarray}%
For a small frequency offset, $\Delta \omega =\omega _{\mathrm{p}}-\omega _{%
	\mathrm{node}}\ll \omega _{\mathrm{p}}$, we have%
\begin{equation}
\cos \left( \frac{\omega _{\mathrm{p}}}{v}L\right) \approx \frac{\pi }{2}%
\frac{\Delta \omega }{\omega _{\mathrm{node}}},
\end{equation}%
with $\cos \left( \omega _{\mathrm{node}}L /v \right) =0$. Then the
Hamiltonian (\ref{Hc}) describes the off-diagonal part of the transmon
Hamiltonian~(\ref{Hamiltonian}) with 
\begin{eqnarray}
G &=&G_{0}\frac{\Delta \omega }{\omega _{\mathrm{node}}},  \notag \\
G_{0}(V_{+}) &=&\frac{\pi }{\hbar }\frac{C_{\mathrm{c}}}{C_{\Sigma }}\sqrt{%
	\frac{\hbar \omega _{10}}{E_{\mathrm{C}}}}eV_{+}\text{.}
\end{eqnarray}%
This is written in the main text as Eq.~(\ref{Bx}).

Consider next the diagonal part of the transmon Hamiltonian given by the
energy-level splitting in Eq.~(\ref{omega10}).%
 
The flux contains the dc and ac components, $\Phi
=\Phi _{\mathrm{dc}}+\Phi _{\mathrm{ac}}\sin \left( \omega _{\mathrm{pump}%
}t\right) $. Assuming the latter being a small value, we obtain%
\begin{equation}
\hbar \omega _{10}=\hbar \omega _{10}(\Phi _{\mathrm{dc}})+\hbar \delta
\left( \Phi _{\mathrm{ac}}\right) \sin \left( \omega _{\mathrm{pump}%
}t\right) ,
\end{equation}%
where $\delta \left( \Phi _{\mathrm{ac}}\right) \propto \Phi _{\mathrm{ac}}$
is the driving amplitude. This is written in the main text as Eq.~(\ref{Bz}).

\begin{figure*}
	\includegraphics[width=0.7 \linewidth]{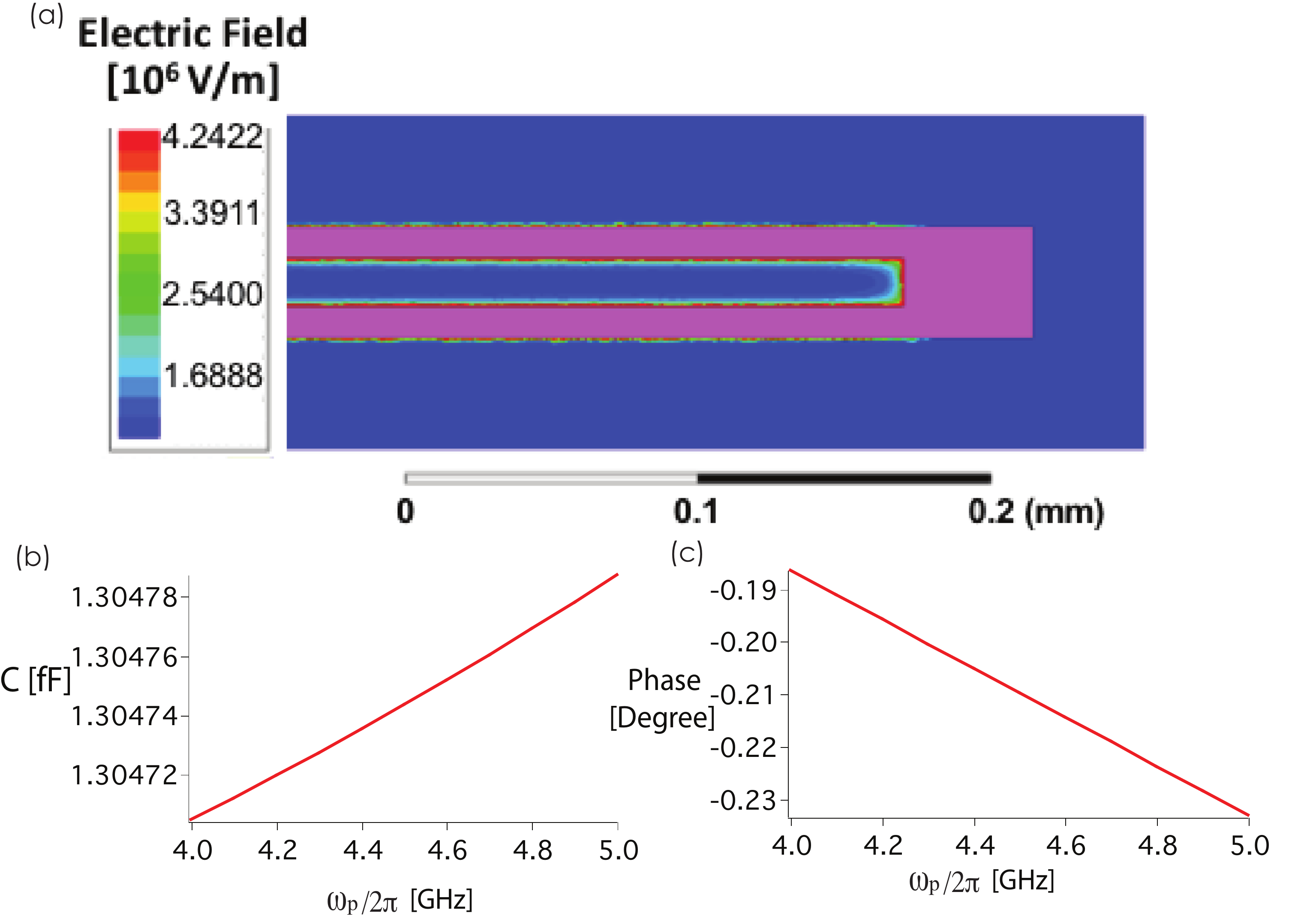}
	\caption{Simulation using High-Frequency Structure Simulator. (a)~Electric field distribution of a transmission line with an end capacitor, acting as a mirror. (b)~The capacitance calculated from the simulated input impedance $Z$, where $Z=1/(\omega C)$. (c)~Phase response of the simulated reflection coefficient caused by the end capacitor.}
	\label{fig:Capacitor}
\end{figure*}

\section{Details of the calculations and the role of nonlinearity}

For obtaining theoretical graphs,  Figs.~3,~4, and~5, we solve \eqref{P1} in each point of the 2D plots.
The amplitude $\delta$ was defined from the lowest pump frequency $\omega_{\rm pump}$, where the first resonance lines vanish, as shown with the yellow double arrow in Fig.~\ref{fig:ExpThrComparison}(a). We choose this, because in the experiment, the side-bands become shifted at low $\omega_{\rm pump}$. We assume that the shift is due to the nonlinearity and take this into account as $\delta = \delta (\omega_{\rm pump})$ below.

The relaxation rate $\Gamma_1$ and the dephasing rate $\Gamma_{\phi}$ are estimated from the experimental data in Fig.~\ref{fig:Flux}(b). We find $\Gamma_2$ from the expression $\Gamma_{2}=\Gamma_{1}/2+\Gamma_{\phi}$. The parameter  $G_0$ is the fitting parameter and we take this equal to 0.1~GHz\,$h$.

In Fig.~\ref{fig:ExpThrComparison} we show details of how we calculate the interferograms and compare with the experiments. Figure~\ref{fig:ExpThrComparison}(b) is the same as Fig.~\ref{fig:PumpFrequencyDependent}(f). In figure~\ref{fig:ExpThrComparison}(c) we take into account the nonlinearity. And in Fig.~\ref{fig:ExpThrComparison}(c,d,e) we can see a good agreement between the location and relative depth of the resonances in the experiment (denoted as Expt) and theory (denoted as Thry) for the high $\omega_{\rm pump}$ frequency, but in the area with low $\omega_{\rm pump}$ we can see increasing the space between the side-bands. In the theory, the location of the side-bands with $\omega_{\rm pump}=0$ is defined by the amplitude $\delta$, and these are located at $\omega_{10}\pm\delta$. From this difference we suggested that in the area of low pump frequency, $\omega_{\rm pump}\rightarrow0$, the amplitude $\delta$ is increased. We take such nonlinearity into account and use this dependence for obtaining a better agreement between the theory and the experiments in the low pumping frequency regime. For this, instead of constant $\delta$, we empirically consider

\begin{equation}
\delta^\prime(\omega_{\rm pump})=\delta+\left(1-\omega_{\rm pump}/G_0\right)^8\delta.
\label{B1}
\end{equation}

Then the obtained interferogram, Fig.~\ref{fig:ExpThrComparison}(c) is in agreement with the experimental Fig.~\ref{fig:ExpThrComparison}(a). This agreement can also be seen in Fig.~\ref{fig:ExpThrComparison}(f), where the brown curve is for the modified theory (denoted as Thry mod). 

\section{Microwave Simulation of end capacitor using High-Frequency Structure Simulator}

We use the 3D electromagnetic simulation software, High-Frequency Structure Simulator, to simulate the end capacitance in the transmission line and the phase shift caused by it. The value of the capacitance (mirror) is about 1.3~fF, see Fig.~\ref{fig:Capacitor}(b), and the phase shift caused by the mirror is less than 0.3~degree, see Fig.~\ref{fig:Capacitor}(c). As expected, the phase shift is very close to the phase shift produced by an ideal mirror, which is zero.





\nocite{apsrev41Control} 
\bibliographystyle{apsrev4-1}
\bibliography{references}

\begin{thebibliography}{45}%
\makeatletter
\providecommand \@ifxundefined [1]{%
 \@ifx{#1\undefined}
}%
\providecommand \@ifnum [1]{%
 \ifnum #1\expandafter \@firstoftwo
 \else \expandafter \@secondoftwo
 \fi
}%
\providecommand \@ifx [1]{%
 \ifx #1\expandafter \@firstoftwo
 \else \expandafter \@secondoftwo
 \fi
}%
\providecommand \natexlab [1]{#1}%
\providecommand \enquote  [1]{``#1''}%
\providecommand \bibnamefont  [1]{#1}%
\providecommand \bibfnamefont [1]{#1}%
\providecommand \citenamefont [1]{#1}%
\providecommand \href@noop [0]{\@secondoftwo}%
\providecommand \href [0]{\begingroup \@sanitize@url \@href}%
\providecommand \@href[1]{\@@startlink{#1}\@@href}%
\providecommand \@@href[1]{\endgroup#1\@@endlink}%
\providecommand \@sanitize@url [0]{\catcode `\\12\catcode `\$12\catcode
  `\&12\catcode `\#12\catcode `\^12\catcode `\_12\catcode `\%12\relax}%
\providecommand \@@startlink[1]{}%
\providecommand \@@endlink[0]{}%
\providecommand \url  [0]{\begingroup\@sanitize@url \@url }%
\providecommand \@url [1]{\endgroup\@href {#1}{\urlprefix }}%
\providecommand \urlprefix  [0]{URL }%
\providecommand \Eprint [0]{\href }%
\providecommand \doibase [0]{http://dx.doi.org/}%
\providecommand \selectlanguage [0]{\@gobble}%
\providecommand \bibinfo  [0]{\@secondoftwo}%
\providecommand \bibfield  [0]{\@secondoftwo}%
\providecommand \translation [1]{[#1]}%
\providecommand \BibitemOpen [0]{}%
\providecommand \bibitemStop [0]{}%
\providecommand \bibitemNoStop [0]{.\EOS\space}%
\providecommand \EOS [0]{\spacefactor3000\relax}%
\providecommand \BibitemShut  [1]{\csname bibitem#1\endcsname}%
\let\auto@bib@innerbib\@empty
\bibitem [{\citenamefont {Gu}\ \emph {et~al.}(2017)\citenamefont {Gu},
  \citenamefont {Kockum}, \citenamefont {Miranowicz}, \citenamefont {Liu},\
  and\ \citenamefont {Nori}}]{Gu2017}%
  \BibitemOpen
  \bibfield  {author} {\bibinfo {author} {\bibfnamefont {X.}~\bibnamefont
  {Gu}}, \bibinfo {author} {\bibfnamefont {A.~F.}\ \bibnamefont {Kockum}},
  \bibinfo {author} {\bibfnamefont {A.}~\bibnamefont {Miranowicz}}, \bibinfo
  {author} {\bibfnamefont {Y.-X.}\ \bibnamefont {Liu}}, \ and\ \bibinfo
  {author} {\bibfnamefont {F.}~\bibnamefont {Nori}},\ }\bibfield  {title}
  {\enquote {\bibinfo {title} {Microwave photonics with superconducting quantum
  circuits},}\ }\href {\doibase 10.1016/j.physrep.2017.10.002} {\bibfield
  {journal} {\bibinfo  {journal} {Phys. Rep.}\ }\textbf {\bibinfo {volume}
  {718-719}},\ \bibinfo {pages} {1--102} (\bibinfo {year} {2017})}\BibitemShut
  {NoStop}%
\bibitem [{\citenamefont {Astafiev}\ \emph {et~al.}(2010)\citenamefont
  {Astafiev}, \citenamefont {Zagoskin}, \citenamefont {Abdumalikov},
  \citenamefont {Pashkin}, \citenamefont {Yamamoto}, \citenamefont {Inomata},
  \citenamefont {Nakamura},\ and\ \citenamefont {Tsai}}]{Astafiev2010}%
  \BibitemOpen
  \bibfield  {author} {\bibinfo {author} {\bibfnamefont {O.}~\bibnamefont
  {Astafiev}}, \bibinfo {author} {\bibfnamefont {A.~M.}\ \bibnamefont
  {Zagoskin}}, \bibinfo {author} {\bibfnamefont {A.~A.}\ \bibnamefont
  {Abdumalikov}}, \bibinfo {author} {\bibfnamefont {Y.~A.}\ \bibnamefont
  {Pashkin}}, \bibinfo {author} {\bibfnamefont {T.}~\bibnamefont {Yamamoto}},
  \bibinfo {author} {\bibfnamefont {K.}~\bibnamefont {Inomata}}, \bibinfo
  {author} {\bibfnamefont {Y.}~\bibnamefont {Nakamura}}, \ and\ \bibinfo
  {author} {\bibfnamefont {J.~S.}\ \bibnamefont {Tsai}},\ }\bibfield  {title}
  {\enquote {\bibinfo {title} {Resonance fluorescence of a single artificial
  atom},}\ }\href {\doibase 10.1126/science.1181918} {\bibfield  {journal}
  {\bibinfo  {journal} {Science}\ }\textbf {\bibinfo {volume} {327}},\ \bibinfo
  {pages} {840--843} (\bibinfo {year} {2010})}\BibitemShut {NoStop}%
\bibitem [{\citenamefont {Lalumiere}\ \emph {et~al.}(2013)\citenamefont
  {Lalumiere}, \citenamefont {Sanders}, \citenamefont {van Loo}, \citenamefont
  {Fedorov}, \citenamefont {Wallraff},\ and\ \citenamefont
  {Blais}}]{Lalumiere13}%
  \BibitemOpen
  \bibfield  {author} {\bibinfo {author} {\bibfnamefont {K.}~\bibnamefont
  {Lalumiere}}, \bibinfo {author} {\bibfnamefont {B.~C.}\ \bibnamefont
  {Sanders}}, \bibinfo {author} {\bibfnamefont {A.~F.}\ \bibnamefont {van
  Loo}}, \bibinfo {author} {\bibfnamefont {A.}~\bibnamefont {Fedorov}},
  \bibinfo {author} {\bibfnamefont {A.}~\bibnamefont {Wallraff}}, \ and\
  \bibinfo {author} {\bibfnamefont {A.}~\bibnamefont {Blais}},\ }\bibfield
  {title} {\enquote {\bibinfo {title} {Input-output theory for waveguide {QED}
  with an ensemble of inhomogeneous atoms},}\ }\href {\doibase
  10.1103/PhysRevA.88.043806} {\bibfield  {journal} {\bibinfo  {journal} {Phys.
  Rev. A}\ }\textbf {\bibinfo {volume} {88}},\ \bibinfo {pages} {043806}
  (\bibinfo {year} {2013})}\BibitemShut {NoStop}%
\bibitem [{\citenamefont {Hoi}\ \emph {et~al.}(2015)\citenamefont {Hoi},
  \citenamefont {Kockum}, \citenamefont {Tornberg}, \citenamefont
  {Pourkabirian}, \citenamefont {Johansson}, \citenamefont {Delsing},\ and\
  \citenamefont {Wilson}}]{Hoi2015}%
  \BibitemOpen
  \bibfield  {author} {\bibinfo {author} {\bibfnamefont {I.-C.}\ \bibnamefont
  {Hoi}}, \bibinfo {author} {\bibfnamefont {A.~F.}\ \bibnamefont {Kockum}},
  \bibinfo {author} {\bibfnamefont {L.}~\bibnamefont {Tornberg}}, \bibinfo
  {author} {\bibfnamefont {A.}~\bibnamefont {Pourkabirian}}, \bibinfo {author}
  {\bibfnamefont {G.}~\bibnamefont {Johansson}}, \bibinfo {author}
  {\bibfnamefont {P.}~\bibnamefont {Delsing}}, \ and\ \bibinfo {author}
  {\bibfnamefont {C.~M.}\ \bibnamefont {Wilson}},\ }\bibfield  {title}
  {\enquote {\bibinfo {title} {Probing the quantum vacuum with an artificial
  atom in front of a mirror},}\ }\href {\doibase 10.1038/nphys3484} {\bibfield
  {journal} {\bibinfo  {journal} {Nat. Phys.}\ }\textbf {\bibinfo {volume}
  {11}},\ \bibinfo {pages} {1045--1049} (\bibinfo {year} {2015})}\BibitemShut
  {NoStop}%
\bibitem [{\citenamefont {Mirhosseini}\ \emph {et~al.}(2019)\citenamefont
  {Mirhosseini}, \citenamefont {Kim}, \citenamefont {Zhang}, \citenamefont
  {Sipahigil}, \citenamefont {Dieterle}, \citenamefont {Keller}, \citenamefont
  {Asenjo-Garcia}, \citenamefont {Chang},\ and\ \citenamefont
  {Painter}}]{Mirhosseini2019}%
  \BibitemOpen
  \bibfield  {author} {\bibinfo {author} {\bibfnamefont {M.}~\bibnamefont
  {Mirhosseini}}, \bibinfo {author} {\bibfnamefont {E.}~\bibnamefont {Kim}},
  \bibinfo {author} {\bibfnamefont {X.}~\bibnamefont {Zhang}}, \bibinfo
  {author} {\bibfnamefont {A.}~\bibnamefont {Sipahigil}}, \bibinfo {author}
  {\bibfnamefont {P.~B.}\ \bibnamefont {Dieterle}}, \bibinfo {author}
  {\bibfnamefont {A.~J.}\ \bibnamefont {Keller}}, \bibinfo {author}
  {\bibfnamefont {A.}~\bibnamefont {Asenjo-Garcia}}, \bibinfo {author}
  {\bibfnamefont {D.~E.}\ \bibnamefont {Chang}}, \ and\ \bibinfo {author}
  {\bibfnamefont {O.}~\bibnamefont {Painter}},\ }\bibfield  {title} {\enquote
  {\bibinfo {title} {Cavity quantum electrodynamics with atom-like mirrors},}\
  }\href {\doibase 10.1038/s41586-019-1196-1} {\bibfield  {journal} {\bibinfo
  {journal} {Nature}\ }\textbf {\bibinfo {volume} {569}},\ \bibinfo {pages}
  {692} (\bibinfo {year} {2019})}\BibitemShut {NoStop}%
\bibitem [{\citenamefont {Hoi}\ \emph {et~al.}(2011)\citenamefont {Hoi},
  \citenamefont {Wilson}, \citenamefont {Johansson}, \citenamefont {Palomaki},
  \citenamefont {Peropadre},\ and\ \citenamefont {Delsing}}]{Hoi2011}%
  \BibitemOpen
  \bibfield  {author} {\bibinfo {author} {\bibfnamefont {I.-C.}\ \bibnamefont
  {Hoi}}, \bibinfo {author} {\bibfnamefont {C.~M.}\ \bibnamefont {Wilson}},
  \bibinfo {author} {\bibfnamefont {G.}~\bibnamefont {Johansson}}, \bibinfo
  {author} {\bibfnamefont {T.}~\bibnamefont {Palomaki}}, \bibinfo {author}
  {\bibfnamefont {B.}~\bibnamefont {Peropadre}}, \ and\ \bibinfo {author}
  {\bibfnamefont {P.}~\bibnamefont {Delsing}},\ }\bibfield  {title} {\enquote
  {\bibinfo {title} {Demonstration of a single-photon router in the microwave
  regime},}\ }\href {\doibase 10.1103/physrevlett.107.073601} {\bibfield
  {journal} {\bibinfo  {journal} {Phys. Rev. Lett.}\ }\textbf {\bibinfo
  {volume} {107}},\ \bibinfo {pages} {073601} (\bibinfo {year}
  {2011})}\BibitemShut {NoStop}%
\bibitem [{\citenamefont {Hoi}\ \emph {et~al.}(2012)\citenamefont {Hoi},
  \citenamefont {Palomaki}, \citenamefont {Lindkvist}, \citenamefont
  {Johansson}, \citenamefont {Delsing},\ and\ \citenamefont
  {Wilson}}]{Hoi2012}%
  \BibitemOpen
  \bibfield  {author} {\bibinfo {author} {\bibfnamefont {I.-C.}\ \bibnamefont
  {Hoi}}, \bibinfo {author} {\bibfnamefont {T.}~\bibnamefont {Palomaki}},
  \bibinfo {author} {\bibfnamefont {J.}~\bibnamefont {Lindkvist}}, \bibinfo
  {author} {\bibfnamefont {G.}~\bibnamefont {Johansson}}, \bibinfo {author}
  {\bibfnamefont {P.}~\bibnamefont {Delsing}}, \ and\ \bibinfo {author}
  {\bibfnamefont {C.~M.}\ \bibnamefont {Wilson}},\ }\bibfield  {title}
  {\enquote {\bibinfo {title} {Generation of nonclassical microwave states
  using an artificial atom in 1{D} open space},}\ }\href {\doibase
  10.1103/PhysRevLett.108.263601} {\bibfield  {journal} {\bibinfo  {journal}
  {Phys. Rev. Lett.}\ }\textbf {\bibinfo {volume} {108}},\ \bibinfo {pages}
  {263601} (\bibinfo {year} {2012})}\BibitemShut {NoStop}%
\bibitem [{\citenamefont {Hoi}\ \emph {et~al.}(2013)\citenamefont {Hoi},
  \citenamefont {Kockum}, \citenamefont {Palomaki}, \citenamefont {Stace},
  \citenamefont {Fan}, \citenamefont {Tornberg}, \citenamefont {Sathyamoorthy},
  \citenamefont {Johansson}, \citenamefont {Delsing},\ and\ \citenamefont
  {Wilson}}]{Hoi2013a}%
  \BibitemOpen
  \bibfield  {author} {\bibinfo {author} {\bibfnamefont {I.-C.}\ \bibnamefont
  {Hoi}}, \bibinfo {author} {\bibfnamefont {A.~F.}\ \bibnamefont {Kockum}},
  \bibinfo {author} {\bibfnamefont {T.}~\bibnamefont {Palomaki}}, \bibinfo
  {author} {\bibfnamefont {T.~M.}\ \bibnamefont {Stace}}, \bibinfo {author}
  {\bibfnamefont {B.}~\bibnamefont {Fan}}, \bibinfo {author} {\bibfnamefont
  {L.}~\bibnamefont {Tornberg}}, \bibinfo {author} {\bibfnamefont {S.~R.}\
  \bibnamefont {Sathyamoorthy}}, \bibinfo {author} {\bibfnamefont
  {G.}~\bibnamefont {Johansson}}, \bibinfo {author} {\bibfnamefont
  {P.}~\bibnamefont {Delsing}}, \ and\ \bibinfo {author} {\bibfnamefont
  {C.~M.}\ \bibnamefont {Wilson}},\ }\bibfield  {title} {\enquote {\bibinfo
  {title} {Giant {Cross-Kerr} effect for propagating microwaves induced by an
  artificial atom},}\ }\href {\doibase 10.1103/physrevlett.111.053601}
  {\bibfield  {journal} {\bibinfo  {journal} {Phys. Rev. Lett.}\ }\textbf
  {\bibinfo {volume} {111}},\ \bibinfo {pages} {053601} (\bibinfo {year}
  {2013})}\BibitemShut {NoStop}%
\bibitem [{\citenamefont {Wen}\ \emph {et~al.}(2018)\citenamefont {Wen},
  \citenamefont {Kockum}, \citenamefont {Ian}, \citenamefont {Chen},
  \citenamefont {Nori},\ and\ \citenamefont {Hoi}}]{wen18}%
  \BibitemOpen
  \bibfield  {author} {\bibinfo {author} {\bibfnamefont {P.~Y.}\ \bibnamefont
  {Wen}}, \bibinfo {author} {\bibfnamefont {A.~F.}\ \bibnamefont {Kockum}},
  \bibinfo {author} {\bibfnamefont {H.}~\bibnamefont {Ian}}, \bibinfo {author}
  {\bibfnamefont {J.~C.}\ \bibnamefont {Chen}}, \bibinfo {author}
  {\bibfnamefont {F.}~\bibnamefont {Nori}}, \ and\ \bibinfo {author}
  {\bibfnamefont {I.-C.}\ \bibnamefont {Hoi}},\ }\bibfield  {title} {\enquote
  {\bibinfo {title} {Reflective amplification without population inversion from
  a strongly driven superconducting qubit},}\ }\href {\doibase
  10.1103/physrevlett.120.063603} {\bibfield  {journal} {\bibinfo  {journal}
  {Phys. Rev. Lett.}\ }\textbf {\bibinfo {volume} {120}},\ \bibinfo {pages}
  {063603} (\bibinfo {year} {2018})}\BibitemShut {NoStop}%
\bibitem [{\citenamefont {Wen}\ \emph {et~al.}(2019)\citenamefont {Wen},
  \citenamefont {Lin}, \citenamefont {Kockum}, \citenamefont {Suri},
  \citenamefont {Ian}, \citenamefont {Chen}, \citenamefont {Mao}, \citenamefont
  {Chiu}, \citenamefont {Delsing}, \citenamefont {Nori}, \citenamefont {Lin},\
  and\ \citenamefont {Hoi}}]{Wen2019}%
  \BibitemOpen
  \bibfield  {author} {\bibinfo {author} {\bibfnamefont {P.~Y.}\ \bibnamefont
  {Wen}}, \bibinfo {author} {\bibfnamefont {K.-T.}\ \bibnamefont {Lin}},
  \bibinfo {author} {\bibfnamefont {A.~F.}\ \bibnamefont {Kockum}}, \bibinfo
  {author} {\bibfnamefont {B.}~\bibnamefont {Suri}}, \bibinfo {author}
  {\bibfnamefont {H.}~\bibnamefont {Ian}}, \bibinfo {author} {\bibfnamefont
  {J.~C.}\ \bibnamefont {Chen}}, \bibinfo {author} {\bibfnamefont {S.~Y.}\
  \bibnamefont {Mao}}, \bibinfo {author} {\bibfnamefont {C.~C.}\ \bibnamefont
  {Chiu}}, \bibinfo {author} {\bibfnamefont {P.}~\bibnamefont {Delsing}},
  \bibinfo {author} {\bibfnamefont {F.}~\bibnamefont {Nori}}, \bibinfo {author}
  {\bibfnamefont {G.-D.}\ \bibnamefont {Lin}}, \ and\ \bibinfo {author}
  {\bibfnamefont {I.-C.}\ \bibnamefont {Hoi}},\ }\bibfield  {title} {\enquote
  {\bibinfo {title} {Large collective {L}amb shift of two distant
  superconducting artificial atoms},}\ }\href {\doibase
  10.1103/PhysRevLett.123.233602} {\bibfield  {journal} {\bibinfo  {journal}
  {Phys. Rev. Lett.}\ }\textbf {\bibinfo {volume} {123}},\ \bibinfo {pages}
  {233602} (\bibinfo {year} {2019})}\BibitemShut {NoStop}%
\bibitem [{\citenamefont {Forn-Diaz}\ \emph {et~al.}(2017)\citenamefont
  {Forn-Diaz}, \citenamefont {Garcia-Ripoll}, \citenamefont {Peropadre},
  \citenamefont {Orgiazzi}, \citenamefont {Yurtalan}, \citenamefont
  {Belyansky}, \citenamefont {Wilson},\ and\ \citenamefont
  {Lupascu}}]{FornDiaz2017}%
  \BibitemOpen
  \bibfield  {author} {\bibinfo {author} {\bibfnamefont {P.}~\bibnamefont
  {Forn-Diaz}}, \bibinfo {author} {\bibfnamefont {J.~J.}\ \bibnamefont
  {Garcia-Ripoll}}, \bibinfo {author} {\bibfnamefont {B.}~\bibnamefont
  {Peropadre}}, \bibinfo {author} {\bibfnamefont {J.-L.}\ \bibnamefont
  {Orgiazzi}}, \bibinfo {author} {\bibfnamefont {M.~A.}\ \bibnamefont
  {Yurtalan}}, \bibinfo {author} {\bibfnamefont {R.}~\bibnamefont {Belyansky}},
  \bibinfo {author} {\bibfnamefont {C.~M.}\ \bibnamefont {Wilson}}, \ and\
  \bibinfo {author} {\bibfnamefont {A.}~\bibnamefont {Lupascu}},\ }\bibfield
  {title} {\enquote {\bibinfo {title} {Ultrastrong coupling of a single
  artificial atom to an electromagnetic continuum in the nonperturbative
  regime},}\ }\href {\doibase 10.1038/nphys3905} {\bibfield  {journal}
  {\bibinfo  {journal} {Nat. Phys.}\ }\textbf {\bibinfo {volume} {13}},\
  \bibinfo {pages} {39--43} (\bibinfo {year} {2017})}\BibitemShut {NoStop}%
\bibitem [{\citenamefont {Szombati}\ \emph {et~al.}(2020)\citenamefont
  {Szombati}, \citenamefont {Gomez~Frieiro}, \citenamefont {M\"uller},
  \citenamefont {Jones}, \citenamefont {Jerger},\ and\ \citenamefont
  {Fedorov}}]{Szombati19}%
  \BibitemOpen
  \bibfield  {author} {\bibinfo {author} {\bibfnamefont {D.}~\bibnamefont
  {Szombati}}, \bibinfo {author} {\bibfnamefont {A.}~\bibnamefont
  {Gomez~Frieiro}}, \bibinfo {author} {\bibfnamefont {C.}~\bibnamefont
  {M\"uller}}, \bibinfo {author} {\bibfnamefont {T.}~\bibnamefont {Jones}},
  \bibinfo {author} {\bibfnamefont {M.}~\bibnamefont {Jerger}}, \ and\ \bibinfo
  {author} {\bibfnamefont {A.}~\bibnamefont {Fedorov}},\ }\bibfield  {title}
  {\enquote {\bibinfo {title} {Quantum rifling: Protecting a qubit from
  measurement back action},}\ }\href {\doibase 10.1103/PhysRevLett.124.070401}
  {\bibfield  {journal} {\bibinfo  {journal} {Phys. Rev. Lett.}\ }\textbf
  {\bibinfo {volume} {124}},\ \bibinfo {pages} {070401} (\bibinfo {year}
  {2020})}\BibitemShut {NoStop}%
\bibitem [{\citenamefont {Karpov}\ \emph {et~al.}(2020)\citenamefont {Karpov},
  \citenamefont {Monarkha}, \citenamefont {Szombati}, \citenamefont {Frieiro},
  \citenamefont {Omelyanchouk}, \citenamefont {Il'ichev}, \citenamefont
  {Fedorov},\ and\ \citenamefont {Shevchenko}}]{Karpov2020}%
  \BibitemOpen
  \bibfield  {author} {\bibinfo {author} {\bibfnamefont {D.~S.}\ \bibnamefont
  {Karpov}}, \bibinfo {author} {\bibfnamefont {V.~Y.}\ \bibnamefont
  {Monarkha}}, \bibinfo {author} {\bibfnamefont {D.}~\bibnamefont {Szombati}},
  \bibinfo {author} {\bibfnamefont {A.~G.}\ \bibnamefont {Frieiro}}, \bibinfo
  {author} {\bibfnamefont {A.~N.}\ \bibnamefont {Omelyanchouk}}, \bibinfo
  {author} {\bibfnamefont {E.}~\bibnamefont {Il'ichev}}, \bibinfo {author}
  {\bibfnamefont {A.}~\bibnamefont {Fedorov}}, \ and\ \bibinfo {author}
  {\bibfnamefont {S.~N.}\ \bibnamefont {Shevchenko}},\ }\bibfield  {title}
  {\enquote {\bibinfo {title} {Probabilistic motional averaging},}\ }\href
  {\doibase 10.1140/epjb/e2019-100514-8} {\bibfield  {journal} {\bibinfo
  {journal} {Eur. Phys. J. B}\ }\textbf {\bibinfo {volume} {93}},\ \bibinfo
  {pages} {49} (\bibinfo {year} {2020})}\BibitemShut {NoStop}%
\bibitem [{\citenamefont {Johansson}\ \emph {et~al.}(2009)\citenamefont
  {Johansson}, \citenamefont {Johansson}, \citenamefont {Wilson},\ and\
  \citenamefont {Nori}}]{Johansson2009}%
  \BibitemOpen
  \bibfield  {author} {\bibinfo {author} {\bibfnamefont {J.~R.}\ \bibnamefont
  {Johansson}}, \bibinfo {author} {\bibfnamefont {G.}~\bibnamefont
  {Johansson}}, \bibinfo {author} {\bibfnamefont {C.~M.}\ \bibnamefont
  {Wilson}}, \ and\ \bibinfo {author} {\bibfnamefont {F.}~\bibnamefont
  {Nori}},\ }\bibfield  {title} {\enquote {\bibinfo {title} {Dynamical
  {C}asimir effect in a superconducting coplanar waveguide},}\ }\href {\doibase
  10.1103/PhysRevLett.103.147003} {\bibfield  {journal} {\bibinfo  {journal}
  {Phys. Rev. Lett.}\ }\textbf {\bibinfo {volume} {103}},\ \bibinfo {pages}
  {147003} (\bibinfo {year} {2009})}\BibitemShut {NoStop}%
\bibitem [{\citenamefont {Johansson}\ \emph {et~al.}(2010)\citenamefont
  {Johansson}, \citenamefont {Johansson}, \citenamefont {Wilson},\ and\
  \citenamefont {Nori}}]{Johansson2010}%
  \BibitemOpen
  \bibfield  {author} {\bibinfo {author} {\bibfnamefont {J.~R.}\ \bibnamefont
  {Johansson}}, \bibinfo {author} {\bibfnamefont {G.}~\bibnamefont
  {Johansson}}, \bibinfo {author} {\bibfnamefont {C.~M.}\ \bibnamefont
  {Wilson}}, \ and\ \bibinfo {author} {\bibfnamefont {F.}~\bibnamefont
  {Nori}},\ }\bibfield  {title} {\enquote {\bibinfo {title} {Dynamical
  {C}asimir effect in superconducting microwave circuits},}\ }\href {\doibase
  10.1103/PhysRevA.82.052509} {\bibfield  {journal} {\bibinfo  {journal} {Phys.
  Rev. A}\ }\textbf {\bibinfo {volume} {82}},\ \bibinfo {pages} {052509}
  (\bibinfo {year} {2010})}\BibitemShut {NoStop}%
\bibitem [{\citenamefont {Wilson}\ \emph {et~al.}(2011)\citenamefont {Wilson},
  \citenamefont {Johansson}, \citenamefont {Pourkabirian}, \citenamefont
  {Simoen}, \citenamefont {Johansson}, \citenamefont {Duty}, \citenamefont
  {Nori},\ and\ \citenamefont {Delsing}}]{Wilson2011}%
  \BibitemOpen
  \bibfield  {author} {\bibinfo {author} {\bibfnamefont {C.~M.}\ \bibnamefont
  {Wilson}}, \bibinfo {author} {\bibfnamefont {G.}~\bibnamefont {Johansson}},
  \bibinfo {author} {\bibfnamefont {A.}~\bibnamefont {Pourkabirian}}, \bibinfo
  {author} {\bibfnamefont {M.}~\bibnamefont {Simoen}}, \bibinfo {author}
  {\bibfnamefont {J.~R.}\ \bibnamefont {Johansson}}, \bibinfo {author}
  {\bibfnamefont {T.}~\bibnamefont {Duty}}, \bibinfo {author} {\bibfnamefont
  {F.}~\bibnamefont {Nori}}, \ and\ \bibinfo {author} {\bibfnamefont
  {P.}~\bibnamefont {Delsing}},\ }\bibfield  {title} {\enquote {\bibinfo
  {title} {Observation of the dynamical {C}asimir effect in a superconducting
  circuit},}\ }\href {\doibase 10.1038/nature10561} {\bibfield  {journal}
  {\bibinfo  {journal} {Nature}\ }\textbf {\bibinfo {volume} {479}},\ \bibinfo
  {pages} {376--379} (\bibinfo {year} {2011})}\BibitemShut {NoStop}%
\bibitem [{\citenamefont {L{\"a}hteenm{\"a}ki}\ \emph
  {et~al.}(2013)\citenamefont {L{\"a}hteenm{\"a}ki}, \citenamefont {Paraoanu},
  \citenamefont {Hassel},\ and\ \citenamefont {Hakonen}}]{Lahteenmaki2013}%
  \BibitemOpen
  \bibfield  {author} {\bibinfo {author} {\bibfnamefont {P.}~\bibnamefont
  {L{\"a}hteenm{\"a}ki}}, \bibinfo {author} {\bibfnamefont {G.~S.}\
  \bibnamefont {Paraoanu}}, \bibinfo {author} {\bibfnamefont {J.}~\bibnamefont
  {Hassel}}, \ and\ \bibinfo {author} {\bibfnamefont {P.~J.}\ \bibnamefont
  {Hakonen}},\ }\bibfield  {title} {\enquote {\bibinfo {title} {{Dynamical
  Casimir effect in a Josephson metamaterial}},}\ }\href {\doibase
  10.1073/pnas.1212705110} {\bibfield  {journal} {\bibinfo  {journal} {PNAS}\
  }\textbf {\bibinfo {volume} {110}},\ \bibinfo {pages} {4234--4238} (\bibinfo
  {year} {2013})}\BibitemShut {NoStop}%
\bibitem [{\citenamefont {Oliver}\ \emph {et~al.}(2005)\citenamefont {Oliver},
  \citenamefont {Yu}, \citenamefont {Lee}, \citenamefont {Berggren},
  \citenamefont {Levitov},\ and\ \citenamefont {Orlando}}]{Oliver2005}%
  \BibitemOpen
  \bibfield  {author} {\bibinfo {author} {\bibfnamefont {W.~D.}\ \bibnamefont
  {Oliver}}, \bibinfo {author} {\bibfnamefont {Y.}~\bibnamefont {Yu}}, \bibinfo
  {author} {\bibfnamefont {J.~C.}\ \bibnamefont {Lee}}, \bibinfo {author}
  {\bibfnamefont {K.~K.}\ \bibnamefont {Berggren}}, \bibinfo {author}
  {\bibfnamefont {L.~S.}\ \bibnamefont {Levitov}}, \ and\ \bibinfo {author}
  {\bibfnamefont {T.~P.}\ \bibnamefont {Orlando}},\ }\bibfield  {title}
  {\enquote {\bibinfo {title} {{Mach-Zehnder} interferometry in a strongly
  driven superconducting qubit},}\ }\href {\doibase 10.1126/science.1119678}
  {\bibfield  {journal} {\bibinfo  {journal} {Science}\ }\textbf {\bibinfo
  {volume} {310}},\ \bibinfo {pages} {1653--1657} (\bibinfo {year}
  {2005})}\BibitemShut {NoStop}%
\bibitem [{\citenamefont {Sillanp\"a\"a}\ \emph {et~al.}(2006)\citenamefont
  {Sillanp\"a\"a}, \citenamefont {Lehtinen}, \citenamefont {Paila},
  \citenamefont {Makhlin},\ and\ \citenamefont {Hakonen}}]{Sillanpaa2006}%
  \BibitemOpen
  \bibfield  {author} {\bibinfo {author} {\bibfnamefont {M.}~\bibnamefont
  {Sillanp\"a\"a}}, \bibinfo {author} {\bibfnamefont {T.}~\bibnamefont
  {Lehtinen}}, \bibinfo {author} {\bibfnamefont {A.}~\bibnamefont {Paila}},
  \bibinfo {author} {\bibfnamefont {Y.}~\bibnamefont {Makhlin}}, \ and\
  \bibinfo {author} {\bibfnamefont {P.}~\bibnamefont {Hakonen}},\ }\bibfield
  {title} {\enquote {\bibinfo {title} {Continuous-time monitoring of
  {Landau-Zener} interference in a {C}ooper-pair box},}\ }\href {\doibase
  10.1103/PhysRevLett.96.187002} {\bibfield  {journal} {\bibinfo  {journal}
  {Phys. Rev. Lett.}\ }\textbf {\bibinfo {volume} {96}},\ \bibinfo {pages}
  {187002} (\bibinfo {year} {2006})}\BibitemShut {NoStop}%
\bibitem [{\citenamefont {Shevchenko}\ \emph {et~al.}(2010)\citenamefont
  {Shevchenko}, \citenamefont {Ashhab},\ and\ \citenamefont
  {Nori}}]{Shevchenko2010}%
  \BibitemOpen
  \bibfield  {author} {\bibinfo {author} {\bibfnamefont {S.~N.}\ \bibnamefont
  {Shevchenko}}, \bibinfo {author} {\bibfnamefont {S.}~\bibnamefont {Ashhab}},
  \ and\ \bibinfo {author} {\bibfnamefont {F.}~\bibnamefont {Nori}},\
  }\bibfield  {title} {\enquote {\bibinfo {title}
  {Landau{\textendash}{Z}ener{\textendash}{S}t{\"u}ckelberg interferometry},}\
  }\href {\doibase 10.1016/j.physrep.2010.03.002} {\bibfield  {journal}
  {\bibinfo  {journal} {Phys. Rep.}\ }\textbf {\bibinfo {volume} {492}},\
  \bibinfo {pages} {1--30} (\bibinfo {year} {2010})}\BibitemShut {NoStop}%
\bibitem [{\citenamefont {van Ditzhuijzen}\ \emph {et~al.}(2009)\citenamefont
  {van Ditzhuijzen}, \citenamefont {Tauschinsky},\ and\ \citenamefont {van
  Linden van~den Heuvell}}]{Ditzhuijzen}%
  \BibitemOpen
  \bibfield  {author} {\bibinfo {author} {\bibfnamefont {C.~S.~E.}\
  \bibnamefont {van Ditzhuijzen}}, \bibinfo {author} {\bibfnamefont
  {A.}~\bibnamefont {Tauschinsky}}, \ and\ \bibinfo {author} {\bibfnamefont
  {H.~B.}\ \bibnamefont {van Linden van~den Heuvell}},\ }\bibfield  {title}
  {\enquote {\bibinfo {title} {Observation of {S}t{\"u}ckelberg oscillations in
  dipole-dipole interactions},}\ }\href {\doibase 10.1103/PhysRevA.80.063407}
  {\bibfield  {journal} {\bibinfo  {journal} {Phys. Rev. A}\ }\textbf {\bibinfo
  {volume} {80}},\ \bibinfo {pages} {063407} (\bibinfo {year}
  {2009})}\BibitemShut {NoStop}%
\bibitem [{\citenamefont {Bogan}\ \emph {et~al.}(2018)\citenamefont {Bogan},
  \citenamefont {Studenikin}, \citenamefont {Korkusinski},\ and\ \citenamefont
  {Gaudreau}}]{Bogan}%
  \BibitemOpen
  \bibfield  {author} {\bibinfo {author} {\bibfnamefont {A.}~\bibnamefont
  {Bogan}}, \bibinfo {author} {\bibfnamefont {S.}~\bibnamefont {Studenikin}},
  \bibinfo {author} {\bibfnamefont {M.}~\bibnamefont {Korkusinski}}, \ and\
  \bibinfo {author} {\bibfnamefont {L.}~\bibnamefont {Gaudreau}},\ }\bibfield
  {title} {\enquote {\bibinfo {title}
  {{Landau-Zener-St}{\"u}ckelberg-{M}ajorana interferometry of a single
  hole},}\ }\href {\doibase 10.1103/PhysRevLett.120.207701} {\bibfield
  {journal} {\bibinfo  {journal} {Phys. Rev. Lett.}\ }\textbf {\bibinfo
  {volume} {120}},\ \bibinfo {pages} {207701} (\bibinfo {year}
  {2018})}\BibitemShut {NoStop}%
\bibitem [{\citenamefont {Mi}\ \emph {et~al.}(2018)\citenamefont {Mi},
  \citenamefont {Kohler},\ and\ \citenamefont {Petta}}]{Mi2018}%
  \BibitemOpen
  \bibfield  {author} {\bibinfo {author} {\bibfnamefont {X.}~\bibnamefont
  {Mi}}, \bibinfo {author} {\bibfnamefont {S.}~\bibnamefont {Kohler}}, \ and\
  \bibinfo {author} {\bibfnamefont {J.~R.}\ \bibnamefont {Petta}},\ }\bibfield
  {title} {\enquote {\bibinfo {title} {Landau-{Z}ener interferometry of
  valley-orbit states in {Si/SiGe} double quantum dots},}\ }\href {\doibase
  10.1103/PhysRevB.98.161404} {\bibfield  {journal} {\bibinfo  {journal} {Phys.
  Rev. B}\ }\textbf {\bibinfo {volume} {98}},\ \bibinfo {pages} {161404(R)}
  (\bibinfo {year} {2018})}\BibitemShut {NoStop}%
\bibitem [{\citenamefont {Ono}\ \emph {et~al.}(2019)\citenamefont {Ono},
  \citenamefont {Shevchenko}, \citenamefont {Mori}, \citenamefont {Moriyama},\
  and\ \citenamefont {Nori}}]{Ono19}%
  \BibitemOpen
  \bibfield  {author} {\bibinfo {author} {\bibfnamefont {K.}~\bibnamefont
  {Ono}}, \bibinfo {author} {\bibfnamefont {S.~N.}\ \bibnamefont {Shevchenko}},
  \bibinfo {author} {\bibfnamefont {T.}~\bibnamefont {Mori}}, \bibinfo {author}
  {\bibfnamefont {S.}~\bibnamefont {Moriyama}}, \ and\ \bibinfo {author}
  {\bibfnamefont {F.}~\bibnamefont {Nori}},\ }\bibfield  {title} {\enquote
  {\bibinfo {title} {Quantum interferometry with a $g$-factor-tunable spin
  qubit},}\ }\href {\doibase 10.1103/PhysRevLett.122.207703} {\bibfield
  {journal} {\bibinfo  {journal} {Phys. Rev. Lett.}\ }\textbf {\bibinfo
  {volume} {122}},\ \bibinfo {pages} {207703} (\bibinfo {year}
  {2019})}\BibitemShut {NoStop}%
\bibitem [{\citenamefont {Otxoa}\ \emph {et~al.}(2019)\citenamefont {Otxoa},
  \citenamefont {Chatterjee}, \citenamefont {Shevchenko}, \citenamefont
  {Barraud}, \citenamefont {Nori},\ and\ \citenamefont
  {Gonzalez-Zalba}}]{Otxoa2019}%
  \BibitemOpen
  \bibfield  {author} {\bibinfo {author} {\bibfnamefont {R.~M.}\ \bibnamefont
  {Otxoa}}, \bibinfo {author} {\bibfnamefont {A.}~\bibnamefont {Chatterjee}},
  \bibinfo {author} {\bibfnamefont {S.~N.}\ \bibnamefont {Shevchenko}},
  \bibinfo {author} {\bibfnamefont {S.}~\bibnamefont {Barraud}}, \bibinfo
  {author} {\bibfnamefont {F.}~\bibnamefont {Nori}}, \ and\ \bibinfo {author}
  {\bibfnamefont {M.~F.}\ \bibnamefont {Gonzalez-Zalba}},\ }\bibfield  {title}
  {\enquote {\bibinfo {title} {Quantum interference capacitor based on
  double-passage {Landau-Zener-St\"uckelberg-Majorana} interferometry},}\
  }\href {\doibase 10.1103/PhysRevB.100.205425} {\bibfield  {journal} {\bibinfo
   {journal} {Phys. Rev. B}\ }\textbf {\bibinfo {volume} {100}},\ \bibinfo
  {pages} {205425} (\bibinfo {year} {2019})}\BibitemShut {NoStop}%
\bibitem [{\citenamefont {Li}\ \emph {et~al.}(2013)\citenamefont {Li},
  \citenamefont {Silveri}, \citenamefont {Kumar}, \citenamefont {Pirkkalainen},
  \citenamefont {Veps\"al\"ainen}, \citenamefont {Chien}, \citenamefont
  {Tuorila}, \citenamefont {Sillanp\"a\"a}, \citenamefont {Hakonen},
  \citenamefont {Thuneberg},\ and\ \citenamefont {Paraoanu}}]{Li13}%
  \BibitemOpen
  \bibfield  {author} {\bibinfo {author} {\bibfnamefont {J.}~\bibnamefont
  {Li}}, \bibinfo {author} {\bibfnamefont {M.~P.}\ \bibnamefont {Silveri}},
  \bibinfo {author} {\bibfnamefont {K.~S.}\ \bibnamefont {Kumar}}, \bibinfo
  {author} {\bibfnamefont {J.-M.}\ \bibnamefont {Pirkkalainen}}, \bibinfo
  {author} {\bibfnamefont {A.}~\bibnamefont {Veps\"al\"ainen}}, \bibinfo
  {author} {\bibfnamefont {W.~C.}\ \bibnamefont {Chien}}, \bibinfo {author}
  {\bibfnamefont {J.}~\bibnamefont {Tuorila}}, \bibinfo {author} {\bibfnamefont
  {M.~A.}\ \bibnamefont {Sillanp\"a\"a}}, \bibinfo {author} {\bibfnamefont
  {P.~J.}\ \bibnamefont {Hakonen}}, \bibinfo {author} {\bibfnamefont {E.~V.}\
  \bibnamefont {Thuneberg}}, \ and\ \bibinfo {author} {\bibfnamefont {G.~S.}\
  \bibnamefont {Paraoanu}},\ }\bibfield  {title} {\enquote {\bibinfo {title}
  {Motional averaging in a superconducting qubit},}\ }\href {\doibase
  10.1038/ncomms2383} {\bibfield  {journal} {\bibinfo  {journal} {Nat. Comm.}\
  }\textbf {\bibinfo {volume} {4}},\ \bibinfo {pages} {1420} (\bibinfo {year}
  {2013})}\BibitemShut {NoStop}%
\bibitem [{\citenamefont {Silveri}\ \emph {et~al.}(2015)\citenamefont
  {Silveri}, \citenamefont {Kumar}, \citenamefont {Tuorila}, \citenamefont
  {Li}, \citenamefont {Vepsalainen}, \citenamefont {Thuneberg},\ and\
  \citenamefont {Paraoanu}}]{Silveri15}%
  \BibitemOpen
  \bibfield  {author} {\bibinfo {author} {\bibfnamefont {M.}~\bibnamefont
  {Silveri}}, \bibinfo {author} {\bibfnamefont {K.}~\bibnamefont {Kumar}},
  \bibinfo {author} {\bibfnamefont {J.}~\bibnamefont {Tuorila}}, \bibinfo
  {author} {\bibfnamefont {J.}~\bibnamefont {Li}}, \bibinfo {author}
  {\bibfnamefont {A.}~\bibnamefont {Vepsalainen}}, \bibinfo {author}
  {\bibfnamefont {E.}~\bibnamefont {Thuneberg}}, \ and\ \bibinfo {author}
  {\bibfnamefont {G.}~\bibnamefont {Paraoanu}},\ }\bibfield  {title} {\enquote
  {\bibinfo {title} {Stueckelberg interference in a superconducting qubit under
  periodic latching modulation},}\ }\href {\doibase
  10.1088/1367-2630/17/4/043058} {\bibfield  {journal} {\bibinfo  {journal}
  {New J. Phys.}\ }\textbf {\bibinfo {volume} {17}},\ \bibinfo {pages} {043058}
  (\bibinfo {year} {2015})}\BibitemShut {NoStop}%
\bibitem [{\citenamefont {Pan}\ \emph {et~al.}(2017)\citenamefont {Pan},
  \citenamefont {Fan}, \citenamefont {Li}, \citenamefont {Dai}, \citenamefont
  {Wei}, \citenamefont {Lu}, \citenamefont {Cao}, \citenamefont {Kang},
  \citenamefont {Xu}, \citenamefont {Chen}, \citenamefont {Sun},\ and\
  \citenamefont {Wu}}]{Pan17}%
  \BibitemOpen
  \bibfield  {author} {\bibinfo {author} {\bibfnamefont {J.}~\bibnamefont
  {Pan}}, \bibinfo {author} {\bibfnamefont {Y.}~\bibnamefont {Fan}}, \bibinfo
  {author} {\bibfnamefont {Y.}~\bibnamefont {Li}}, \bibinfo {author}
  {\bibfnamefont {X.}~\bibnamefont {Dai}}, \bibinfo {author} {\bibfnamefont
  {X.}~\bibnamefont {Wei}}, \bibinfo {author} {\bibfnamefont {Y.}~\bibnamefont
  {Lu}}, \bibinfo {author} {\bibfnamefont {C.}~\bibnamefont {Cao}}, \bibinfo
  {author} {\bibfnamefont {L.}~\bibnamefont {Kang}}, \bibinfo {author}
  {\bibfnamefont {W.}~\bibnamefont {Xu}}, \bibinfo {author} {\bibfnamefont
  {J.}~\bibnamefont {Chen}}, \bibinfo {author} {\bibfnamefont {G.}~\bibnamefont
  {Sun}}, \ and\ \bibinfo {author} {\bibfnamefont {P.}~\bibnamefont {Wu}},\
  }\bibfield  {title} {\enquote {\bibinfo {title} {Dynamically modulated
  {Autler-Townes} effect in a transmon qubit},}\ }\href {\doibase
  10.1103/PhysRevB.96.024502} {\bibfield  {journal} {\bibinfo  {journal} {Phys.
  Rev. B}\ }\textbf {\bibinfo {volume} {96}},\ \bibinfo {pages} {024502}
  (\bibinfo {year} {2017})}\BibitemShut {NoStop}%
\bibitem [{\citenamefont {Bera}\ \emph {et~al.}(2020)\citenamefont {Bera},
  \citenamefont {Majumder}, \citenamefont {Sahu},\ and\ \citenamefont
  {Singh}}]{Bera2020}%
  \BibitemOpen
  \bibfield  {author} {\bibinfo {author} {\bibfnamefont {T.}~\bibnamefont
  {Bera}}, \bibinfo {author} {\bibfnamefont {S.}~\bibnamefont {Majumder}},
  \bibinfo {author} {\bibfnamefont {S.~K.}\ \bibnamefont {Sahu}}, \ and\
  \bibinfo {author} {\bibfnamefont {V.}~\bibnamefont {Singh}},\ }\bibfield
  {title} {\enquote {\bibinfo {title} {Large flux-mediated coupling in hybrid
  electromechanical system with a transmon qubit},}\ }\href@noop {} {\bibfield
  {journal} {\bibinfo  {journal} {arXiv:2001.05700}\ } (\bibinfo {year}
  {2020})}\BibitemShut {NoStop}%
\bibitem [{\citenamefont {Eschner}\ \emph {et~al.}(2001)\citenamefont
  {Eschner}, \citenamefont {Raab}, \citenamefont {Schmidt-Kaler},\ and\
  \citenamefont {Blatt}}]{Eschner}%
  \BibitemOpen
  \bibfield  {author} {\bibinfo {author} {\bibfnamefont {J.}~\bibnamefont
  {Eschner}}, \bibinfo {author} {\bibfnamefont {C.}~\bibnamefont {Raab}},
  \bibinfo {author} {\bibfnamefont {F.}~\bibnamefont {Schmidt-Kaler}}, \ and\
  \bibinfo {author} {\bibfnamefont {R.}~\bibnamefont {Blatt}},\ }\bibfield
  {title} {\enquote {\bibinfo {title} {Light interference from single atoms and
  their mirror images},}\ }\href {\doibase 10.1038/35097017} {\bibfield
  {journal} {\bibinfo  {journal} {Nature}\ }\textbf {\bibinfo {volume} {413}},\
  \bibinfo {pages} {495} (\bibinfo {year} {2001})}\BibitemShut {NoStop}%
\bibitem [{\citenamefont {Gorelik}\ \emph {et~al.}(1998)\citenamefont
  {Gorelik}, \citenamefont {Lundin}, \citenamefont {Shumeiko}, \citenamefont
  {Shekhter},\ and\ \citenamefont {Jonson}}]{Gorelik1998}%
  \BibitemOpen
  \bibfield  {author} {\bibinfo {author} {\bibfnamefont {L.~Y.}\ \bibnamefont
  {Gorelik}}, \bibinfo {author} {\bibfnamefont {N.~I.}\ \bibnamefont {Lundin}},
  \bibinfo {author} {\bibfnamefont {V.~S.}\ \bibnamefont {Shumeiko}}, \bibinfo
  {author} {\bibfnamefont {R.~I.}\ \bibnamefont {Shekhter}}, \ and\ \bibinfo
  {author} {\bibfnamefont {M.}~\bibnamefont {Jonson}},\ }\bibfield  {title}
  {\enquote {\bibinfo {title} {Superconducting single-mode contact as a
  microwave-activated quantum interferometer},}\ }\href {\doibase
  10.1103/physrevlett.81.2538} {\bibfield  {journal} {\bibinfo  {journal}
  {Phys. Rev. Lett.}\ }\textbf {\bibinfo {volume} {81}},\ \bibinfo {pages}
  {2538--2541} (\bibinfo {year} {1998})}\BibitemShut {NoStop}%
\bibitem [{\citenamefont {Wu}\ \emph {et~al.}(2019)\citenamefont {Wu},
  \citenamefont {Zhou}, \citenamefont {Xu}, \citenamefont {Liu},\ and\
  \citenamefont {Li}}]{Wu19}%
  \BibitemOpen
  \bibfield  {author} {\bibinfo {author} {\bibfnamefont {T.}~\bibnamefont
  {Wu}}, \bibinfo {author} {\bibfnamefont {Y.}~\bibnamefont {Zhou}}, \bibinfo
  {author} {\bibfnamefont {Y.}~\bibnamefont {Xu}}, \bibinfo {author}
  {\bibfnamefont {S.}~\bibnamefont {Liu}}, \ and\ \bibinfo {author}
  {\bibfnamefont {J.}~\bibnamefont {Li}},\ }\bibfield  {title} {\enquote
  {\bibinfo {title} {{Landau-Zener-S}t\"uckelberg interference in nonlinear
  regime},}\ }\href {\doibase 10.1088/0256-307X/36/12/124204} {\bibfield
  {journal} {\bibinfo  {journal} {Chin. Phys. Lett.}\ }\textbf {\bibinfo
  {volume} {36}},\ \bibinfo {pages} {124204} (\bibinfo {year}
  {2019})}\BibitemShut {NoStop}%
\bibitem [{\citenamefont {Gong}\ \emph {et~al.}(2016)\citenamefont {Gong},
  \citenamefont {Zhou}, \citenamefont {Lan}, \citenamefont {Fan}, \citenamefont
  {Pan}, \citenamefont {Yu}, \citenamefont {Chen}, \citenamefont {Sun},
  \citenamefont {Yu}, \citenamefont {Han},\ and\ \citenamefont {Wu}}]{Gong16}%
  \BibitemOpen
  \bibfield  {author} {\bibinfo {author} {\bibfnamefont {M.}~\bibnamefont
  {Gong}}, \bibinfo {author} {\bibfnamefont {Y.}~\bibnamefont {Zhou}}, \bibinfo
  {author} {\bibfnamefont {D.}~\bibnamefont {Lan}}, \bibinfo {author}
  {\bibfnamefont {Y.}~\bibnamefont {Fan}}, \bibinfo {author} {\bibfnamefont
  {J.}~\bibnamefont {Pan}}, \bibinfo {author} {\bibfnamefont {H.}~\bibnamefont
  {Yu}}, \bibinfo {author} {\bibfnamefont {J.}~\bibnamefont {Chen}}, \bibinfo
  {author} {\bibfnamefont {G.}~\bibnamefont {Sun}}, \bibinfo {author}
  {\bibfnamefont {Y.}~\bibnamefont {Yu}}, \bibinfo {author} {\bibfnamefont
  {S.}~\bibnamefont {Han}}, \ and\ \bibinfo {author} {\bibfnamefont
  {P.}~\bibnamefont {Wu}},\ }\bibfield  {title} {\enquote {\bibinfo {title}
  {{Landau-Zener-St\"uckelberg-Majorana} interference in a {3D} transmon driven
  by a chirped microwave},}\ }\href {\doibase 10.1063/1.4944327} {\bibfield
  {journal} {\bibinfo  {journal} {Applied Phys. Lett.}\ }\textbf {\bibinfo
  {volume} {108}},\ \bibinfo {eid} {112602} (\bibinfo {year}
  {2016})}\BibitemShut {NoStop}%
\bibitem [{\citenamefont {Satanin}\ \emph {et~al.}(2014)\citenamefont
  {Satanin}, \citenamefont {Denisenko}, \citenamefont {Gelman},\ and\
  \citenamefont {Nori}}]{Satanin2014}%
  \BibitemOpen
  \bibfield  {author} {\bibinfo {author} {\bibfnamefont {A.~M.}\ \bibnamefont
  {Satanin}}, \bibinfo {author} {\bibfnamefont {M.~V.}\ \bibnamefont
  {Denisenko}}, \bibinfo {author} {\bibfnamefont {A.~I.}\ \bibnamefont
  {Gelman}}, \ and\ \bibinfo {author} {\bibfnamefont {F.}~\bibnamefont
  {Nori}},\ }\bibfield  {title} {\enquote {\bibinfo {title} {Amplitude and
  phase effects in {J}osephson qubits driven by a biharmonic electromagnetic
  field},}\ }\href {\doibase 10.1103/physrevb.90.104516} {\bibfield  {journal}
  {\bibinfo  {journal} {Phys. Rev. B}\ }\textbf {\bibinfo {volume} {90}},\
  \bibinfo {pages} {104516} (\bibinfo {year} {2014})}\BibitemShut {NoStop}%
\bibitem [{\citenamefont {Giavaras}\ and\ \citenamefont
  {Tokura}(2019)}]{Giavaras2020}%
  \BibitemOpen
  \bibfield  {author} {\bibinfo {author} {\bibfnamefont {G.}~\bibnamefont
  {Giavaras}}\ and\ \bibinfo {author} {\bibfnamefont {Y.}~\bibnamefont
  {Tokura}},\ }\bibfield  {title} {\enquote {\bibinfo {title} {Spectroscopy of
  double quantum dot two-spin states by tuning the interdot barrier},}\ }\href
  {\doibase 10.1103/PhysRevB.99.075412} {\bibfield  {journal} {\bibinfo
  {journal} {Phys. Rev. B}\ }\textbf {\bibinfo {volume} {99}},\ \bibinfo
  {pages} {075412} (\bibinfo {year} {2019})}\BibitemShut {NoStop}%
\bibitem [{\citenamefont {You}\ \emph {et~al.}(2006)\citenamefont {You},
  \citenamefont {Tsai},\ and\ \citenamefont {Nori}}]{You2006}%
  \BibitemOpen
  \bibfield  {author} {\bibinfo {author} {\bibfnamefont {J.~Q.}\ \bibnamefont
  {You}}, \bibinfo {author} {\bibfnamefont {J.~S.}\ \bibnamefont {Tsai}}, \
  and\ \bibinfo {author} {\bibfnamefont {F.}~\bibnamefont {Nori}},\ }\bibfield
  {title} {\enquote {\bibinfo {title} {Hybridized solid-state qubit in the
  charge-flux regime},}\ }\href {\doibase 10.1103/physrevb.73.014510}
  {\bibfield  {journal} {\bibinfo  {journal} {Phys. Rev. B}\ }\textbf {\bibinfo
  {volume} {73}},\ \bibinfo {pages} {014510} (\bibinfo {year}
  {2006})}\BibitemShut {NoStop}%
\bibitem [{\citenamefont {You}\ \emph {et~al.}(2007)\citenamefont {You},
  \citenamefont {Hu}, \citenamefont {Ashhab},\ and\ \citenamefont
  {Nori}}]{You2007}%
  \BibitemOpen
  \bibfield  {author} {\bibinfo {author} {\bibfnamefont {J.~Q.}\ \bibnamefont
  {You}}, \bibinfo {author} {\bibfnamefont {X.}~\bibnamefont {Hu}}, \bibinfo
  {author} {\bibfnamefont {S.}~\bibnamefont {Ashhab}}, \ and\ \bibinfo {author}
  {\bibfnamefont {F.}~\bibnamefont {Nori}},\ }\bibfield  {title} {\enquote
  {\bibinfo {title} {Low-decoherence flux qubit},}\ }\href {\doibase
  10.1103/physrevb.75.140515} {\bibfield  {journal} {\bibinfo  {journal} {Phys.
  Rev. B}\ }\textbf {\bibinfo {volume} {75}},\ \bibinfo {pages} {140515}
  (\bibinfo {year} {2007})}\BibitemShut {NoStop}%
\bibitem [{\citenamefont {Kockum}\ and\ \citenamefont
  {Nori}(2019)}]{Kockum2019}%
  \BibitemOpen
  \bibfield  {author} {\bibinfo {author} {\bibfnamefont {A.~F.}\ \bibnamefont
  {Kockum}}\ and\ \bibinfo {author} {\bibfnamefont {F.}~\bibnamefont {Nori}},\
  }\bibfield  {title} {\enquote {\bibinfo {title} {{Quantum Bits with Josephson
  Junctions}},}\ }in\ \href {\doibase 10.1007/978-3-030-20726-7_17} {\emph
  {\bibinfo {booktitle} {Fundamentals and Frontiers of the Josephson Effect}}}\
  (\bibinfo  {publisher} {Springer International Publishing},\ \bibinfo {year}
  {2019})\ pp.\ \bibinfo {pages} {703--741}\BibitemShut {NoStop}%
\bibitem [{\citenamefont {Ono}\ \emph {et~al.}(2020)\citenamefont {Ono},
  \citenamefont {Shevchenko}, \citenamefont {Mori}, \citenamefont {Moriyama},\
  and\ \citenamefont {Nori}}]{Ono2020}%
  \BibitemOpen
  \bibfield  {author} {\bibinfo {author} {\bibfnamefont {K.}~\bibnamefont
  {Ono}}, \bibinfo {author} {\bibfnamefont {S.~N.}\ \bibnamefont {Shevchenko}},
  \bibinfo {author} {\bibfnamefont {T.}~\bibnamefont {Mori}}, \bibinfo {author}
  {\bibfnamefont {S.}~\bibnamefont {Moriyama}}, \ and\ \bibinfo {author}
  {\bibfnamefont {F.}~\bibnamefont {Nori}},\ }\bibfield  {title} {\enquote
  {\bibinfo {title} {Single-spin qubit analogous to a quantum heat engine},}\
  }\href@noop {} {\bibfield  {journal} {\bibinfo  {journal} {arXiv:2008.10181}\
  } (\bibinfo {year} {2020})}\BibitemShut {NoStop}%
\bibitem [{\citenamefont {Silveri}\ \emph {et~al.}(2017)\citenamefont
  {Silveri}, \citenamefont {Tuorila}, \citenamefont {Thuneberg},\ and\
  \citenamefont {Paraoanu}}]{Silveri2017}%
  \BibitemOpen
  \bibfield  {author} {\bibinfo {author} {\bibfnamefont {M.~P.}\ \bibnamefont
  {Silveri}}, \bibinfo {author} {\bibfnamefont {J.~A.}\ \bibnamefont
  {Tuorila}}, \bibinfo {author} {\bibfnamefont {E.~V.}\ \bibnamefont
  {Thuneberg}}, \ and\ \bibinfo {author} {\bibfnamefont {G.~S.}\ \bibnamefont
  {Paraoanu}},\ }\bibfield  {title} {\enquote {\bibinfo {title} {Quantum
  systems under frequency modulation},}\ }\href {\doibase
  10.1088/1361-6633/aa5170} {\bibfield  {journal} {\bibinfo  {journal} {Rep.
  Prog. Phys.}\ }\textbf {\bibinfo {volume} {80}},\ \bibinfo {pages} {056002}
  (\bibinfo {year} {2017})}\BibitemShut {NoStop}%
\bibitem [{\citenamefont {Ivakhnenko}\ \emph {et~al.}(2018)\citenamefont
  {Ivakhnenko}, \citenamefont {Shevchenko},\ and\ \citenamefont
  {Nori}}]{Ivakhnenko18}%
  \BibitemOpen
  \bibfield  {author} {\bibinfo {author} {\bibfnamefont {O.~V.}\ \bibnamefont
  {Ivakhnenko}}, \bibinfo {author} {\bibfnamefont {S.~N.}\ \bibnamefont
  {Shevchenko}}, \ and\ \bibinfo {author} {\bibfnamefont {F.}~\bibnamefont
  {Nori}},\ }\bibfield  {title} {\enquote {\bibinfo {title} {Simulating quantum
  dynamical phenomena using classical oscillators:
  {Landau-Zener-St\"uckelberg-Majorana} interferometry, latching modulation,
  and motional averaging},}\ }\href {\doibase 10.1038/s41598-018-28993-8}
  {\bibfield  {journal} {\bibinfo  {journal} {Sci. Rep.}\ }\textbf {\bibinfo
  {volume} {8}},\ \bibinfo {pages} {12218} (\bibinfo {year}
  {2018})}\BibitemShut {NoStop}%
\bibitem [{\citenamefont {Hoi}(2013)}]{HoiPhD}%
  \BibitemOpen
  \bibfield  {author} {\bibinfo {author} {\bibfnamefont {I.-C.}\ \bibnamefont
  {Hoi}},\ }\emph {\bibinfo {title} {Quantum Optics with Propagating Microwaves
  in Superconducting Circuits}},\ \href@noop {} {Ph.D. thesis},\ \bibinfo
  {school} {Chalmers University of Technology, Sweden}, \bibinfo {address}
  {Gothenburg, Sweden} (\bibinfo {year} {2013})\BibitemShut {NoStop}%
\bibitem [{\citenamefont {Wendin}(2017)}]{Wendin2017}%
  \BibitemOpen
  \bibfield  {author} {\bibinfo {author} {\bibfnamefont {G.}~\bibnamefont
  {Wendin}},\ }\bibfield  {title} {\enquote {\bibinfo {title} {Quantum
  information processing with superconducting circuits: a review},}\ }\href
  {\doibase 10.1088/1361-6633/aa7e1a} {\bibfield  {journal} {\bibinfo
  {journal} {Rep. Progr. Phys.}\ }\textbf {\bibinfo {volume} {80}},\ \bibinfo
  {pages} {106001} (\bibinfo {year} {2017})}\BibitemShut {NoStop}%
\bibitem [{\citenamefont {Koch}\ \emph {et~al.}(2007)\citenamefont {Koch},
  \citenamefont {Yu}, \citenamefont {Gambetta}, \citenamefont {Houck},
  \citenamefont {Schuster}, \citenamefont {Majer}, \citenamefont {Blais},
  \citenamefont {Devoret}, \citenamefont {Girvin},\ and\ \citenamefont
  {Schoelkopf}}]{Koch2007}%
  \BibitemOpen
  \bibfield  {author} {\bibinfo {author} {\bibfnamefont {J.}~\bibnamefont
  {Koch}}, \bibinfo {author} {\bibfnamefont {T.~M.}\ \bibnamefont {Yu}},
  \bibinfo {author} {\bibfnamefont {J.}~\bibnamefont {Gambetta}}, \bibinfo
  {author} {\bibfnamefont {A.~A.}\ \bibnamefont {Houck}}, \bibinfo {author}
  {\bibfnamefont {D.~I.}\ \bibnamefont {Schuster}}, \bibinfo {author}
  {\bibfnamefont {J.}~\bibnamefont {Majer}}, \bibinfo {author} {\bibfnamefont
  {A.}~\bibnamefont {Blais}}, \bibinfo {author} {\bibfnamefont {M.~H.}\
  \bibnamefont {Devoret}}, \bibinfo {author} {\bibfnamefont {S.~M.}\
  \bibnamefont {Girvin}}, \ and\ \bibinfo {author} {\bibfnamefont {R.~J.}\
  \bibnamefont {Schoelkopf}},\ }\bibfield  {title} {\enquote {\bibinfo {title}
  {Charge-insensitive qubit design derived from the {Cooper} pair box},}\
  }\href {\doibase 10.1103/physreva.76.042319} {\bibfield  {journal} {\bibinfo
  {journal} {Phys. Rev. A}\ }\textbf {\bibinfo {volume} {76}},\ \bibinfo
  {pages} {042319} (\bibinfo {year} {2007})}\BibitemShut {NoStop}%
\bibitem [{\citenamefont {Shevchenko}\ and\ \citenamefont
  {Karpov}(2018)}]{Shevchenko18}%
  \BibitemOpen
  \bibfield  {author} {\bibinfo {author} {\bibfnamefont {S.~N.}\ \bibnamefont
  {Shevchenko}}\ and\ \bibinfo {author} {\bibfnamefont {D.~S.}\ \bibnamefont
  {Karpov}},\ }\bibfield  {title} {\enquote {\bibinfo {title} {Thermometry and
  memcapacitance with qubit-resonator system},}\ }\href {\doibase
  10.1103/PhysRevApplied.10.014013} {\bibfield  {journal} {\bibinfo  {journal}
  {Phys. Rev. Applied}\ }\textbf {\bibinfo {volume} {10}},\ \bibinfo {pages}
  {014013} (\bibinfo {year} {2018})}\BibitemShut {NoStop}%
\end{thebibliography}%

\end{document}